# Experimental studies of crystal nucleation: metals and colloids

Dieter M. Herlach[1], Thomas Palberg[2], Ina Klassen[1)3)], Stefan Klein[1], Raphael Kobold [1]

[1] *Institut für Materialphysik im Weltraum, Deutsches Zentrum für Luft- und Raumfahrt (DLR), 51170 Köln, Germany*
[2] *Institut für Physik, Johannes Gutenberg Universität Mainz, 55099 Mainz, Germany*
[3] *Projektträger Jülich, Forschungszentrum Jülich GmbH, 52425 Jülich, Germany*

Crystallization is one of the most important phase transformations of first order. In the case of metals and alloys, the liquid phase is the parent phase of materials production. The conditions of the crystallization process control the as-solidified material in its chemical and physical properties. Nucleation initiates the crystallization of a liquid. It selects the crystallographic phase, stable or meta-stable. Its detailed knowledge is therefore mandatory for the design of materials. We present techniques of containerless processing for nucleation studies of metals and alloys. Experimental results demonstrate the power of these methods not only for crystal nucleation of stable solids but in particular also for investigations of crystal nucleation of metastable solids at extreme undercooling. This concerns the physical nature of heterogeneous versus homogeneous nucleation and nucleation of phases nucleated under non-equilibrium conditions. The results are analyzed within classical nucleation theory that defines the activation energy of homogeneous nucleation in terms of the interfacial energy and the difference of Gibbs free energies of solid and liquid. The interfacial energy acts as barrier for the nucleation process. Its experimental determination is difficult in the case of metals. In the second part of this work we therefore explore the potential of colloidal suspensions as model systems for the crystallization process. The nucleation process of colloids is observed in situ by optical observation and ultra-small angle X-ray diffraction using high intensity synchrotron radiation. It allows an unambiguous discrimination of homogeneous and heterogeneous nucleation as well as the determination of the interfacial free energy of the solid-liquid interface. Our results are used to construct Turnbull plots of colloids, which are discussed in relation to Turnbull plots of metals and support the hypothesis that colloids are useful model systems to investigate crystal nucleation.

# 1 Introduction

Metallic materials are prepared from the liquid state as their parent phase. The conditions under which the liquid solidifies determine the physical and chemical properties of the as-solidified material. In most cases, time and energy consuming post-solidification treatment of the ingot material is mandatory to obtain the final product with its desired properties and design performance. Therefore, efforts are directed towards virtual material design with computer assisted modelling. This can shorten the entire production chain - ranging from casting the shaped solid from the melt to the final tuning of the product in order to save costs during the production process. The goal is to fabricate novel materials with improved properties for specific applications.

Computational materials science performed from the liquid state requires detailed knowledge of the physical mechanisms involved in the solidification process. In particular, crystal nucleation is the decisive process that initiates solidification and pre-selects the crystallographic phase either stable or metastable. The short-range order in the liquid as precursor often influences nucleation and the phase transformations of first order. Nucleation is a thermally activated process that requires an undercooling of the liquid below its equilibrium melting temperature to create a driving force for the formation of supercritical nuclei. In particular, at large undercooling, nucleation pathways for different crystallographic phases can compete with each other. This gives access to non-equilibrium solidification, which can form meta-stable solids, which may differ in their physical and chemical properties from their stable counterparts. Detailed modelling of solidification, far away from thermodynamic equilibrium, requires that the solidification process must be investigated in every detail [1].

In order to achieve the state of an undercooled melt, it is advantageous to remove heterogeneous nucleation sites which otherwise limit the amount of achievable undercooling. One efficient way to realize such conditions is containerless processing of melts [2]. In such, the most dominant heterogeneous nucleation process, involving interaction with container walls, is completely avoided. Nowadays, electromagnetic [3] and electrostatic levitation techniques [4, 5] have been developed for containerless undercooling and solidification of molten metals and alloys. A freely suspended drop gives the extra benefit to directly observe the solidification process by combining the levitation technique with proper diagnostic means [6, 7]. Short range ordering in undercooled metallic melts as precursor of crystal nucleation has been investigated by using neutron diffraction [8] and synchrotron radiation [9] on containerless undercooled melts [10]. Additionally, primary phase selection processes for rapid solidification of meta-stable phases has been observed in situ by energy dispersive X-ray radiation using synchrotron radiation of high intensity [11].

Nucleation is governed by heat and mass transport. Thus, any transport process stimulated externally by natural convection and/or forced convection due to stirring effects of alternating electromagnetic fields in electromagnetic levitation experiments may cause serious influence on the solidification process. To understand this effect and to develop a quantitative description of crystallization in the presence of forced convection comparative experiments on Earth and in reduced gravity are of great help. Under the special conditions of reduced gravity for instance in Space, the forces needed to compensate disturbing accelerations are about three orders of magnitude smaller than the forces needed to compensate the gravitational force for levitation experiments on Earth. In a cooperative effort by the European Space Agency (ESA) and the German Space Center – Space Management (DLR), a facility for Electro-Magnetic Levitation (EML) was constructed and accommodated on board the *Columbus* European module of the International Space Station (ISS). Many experiments are in progress by international investigator teams to apply this multi-user facility for investigations on undercooled melts under the specific environment conditions of reduced gravity in Space [12].

Even though a great variety of models ranging from physical atomistic to pure phenomenological models have been developed to understand and even to quantitatively describe nucleation processes, experimental results do not cover the entire spectrum of phenomena involved in nucleation [13]. The direct observation of nucleation process in atomic and molecular systems suffers from the fact that the materials of interest are often non transparent and relaxation of atoms or molecules is very fast. Their vibration frequencies range up to $10^{13}$ Hertz comparable with the Debye frequency. This makes direct observation of nucleation in undercooled metals extremely difficult or even impossible.

These limitations are overcome if model systems are investigated. Both digital computer simulations and analog experimental simulations using colloids have extensively been applied to study the fundamentals of nucleation. In such approaches, however, it is crucial to obtain detailed insight to what extent such model systems can be directly compared to atomic systems like metals and alloys. As a first step, one may compare results on colloidal crystallization in vivo and in vitro [14]. The decisive step, however, is the direct comparison of experiments on colloidal models to those of metals [15, 16, 17]. This is also attempted in the previous article. This idea to pursue nucleation in parallel in such different systems as metals and colloids in fact originates from the last century. The present paper gives an outline, how far this comparison may be taken.

Colloidal suspensions consist of mesoscopic particles (of about 10 nm – 10 μm in size) dispersed in a carrier fluid like water or oil. A great advantage of these mesoscopic models opposite to atomic or molecular systems, where the interaction potential is fixed, is the possibility of a precise and controlled tuning of the interaction potential at constant temperature [18]. Investigations include hard spheres, charged spheres and entropically attractive systems. Monodisperse colloidal spheres spontaneously form so-called colloidal fluids or solids, if the experimentally variable interaction between the particles is strong enough [19, 20] or the entropy difference between fluid and crystal becomes sufficiently large [21]. Typical structures of these single component systems comprise a strongly correlated fluid [22] as well as body centered cubic (bcc) [23] and face centered cubic (fcc) [24] phases, very analogous to many metals. Mixtures form spindle type, aceotropic or eutectic phase diagrams [25]. Moreover, a large variety of alloys and even Laves phases are known [26, 27]. Further, Brownian motion governs the movement of colloidal particles and limits the time scales of colloidal dynamics to the convenient narrow range [28].

The time scale of Brownian motion is by several orders of magnitude more sluggish compared with the Debye frequency. Within milliseconds a typical interparticle distance is crossed. Nucleation needs seconds to minutes [29], growth needs some minutes to a few hours for completing the phase transformations [14, 30]. Coarsening occurs within a few days at room temperature [31, 32]. These highly correlated systems become widely recognized as model systems for condensed matter physics. Since the nineties of the last century, colloidal suspensions have been viewed as potential models for fundamental investigations on the structure and dynamics of condensed matter [21], and the 'colloids as atoms' paradigm has led to many interesting studies on their phase behavior and phase transition kinetics [30, 33, 34, 35, 36, 37, 38, 39]. They are guided by valuable instrumental and theoretical developments and complemented by a multitude of computer studies [40]. The typical time and length scales of colloidal systems allow for time-resolved observations of solidification with easily manageable experimental techniques, even real space imaging of nucleation events with 'atomic' resolution [41, 42, [43]. More recently, it was demonstrated from scattering experiments that also the thermodynamics of crystallization can be accessed in a quantitative manner and the Turnbull coefficients of bcc forming metals and colloids are very similar [44]. That stresses the applicability of colloidal suspensions as model systems for nucleation studies, which are not directly applicable for metallic systems.

In the present work we will focus on the presentation of experimental techniques to perform nucleation studies in undercooled metals and alloys, and colloidal suspension as well. Exemplary results obtained by these methods are presented which allow for a detailed discussion of nucleation phenomena in metallic and colloidal systems.

## 2 Experimental set ups

### 2.1 Metals

#### 2.1.1 Electromagnetic levitation

For metallic systems the most suitable technique for freely suspending spheres of diameter up to 1cm is the electromagnetic levitation technique. Figure 1 shows an electromagnetically levitated sphere in a levitation coil. The principle of electromagnetic levitation is based on the induction of eddy currents in an electrically conducting material if the material experiences a time dependent magnetic field $B$ (Lenz rule)

$$\nabla \times E = -\partial B / \partial t \tag{2.1}$$

with $E$ the electrostatic field. For a non-uniform magnetic field the eddy currents induced in a sample produce a magnetic dipole moment $m$ that is opposite to the primary field $B$. This leads to a diamagnetic repulsion force $F_r$

$$F_r = -\nabla(m \cdot B) \tag{2.2}$$

between the primary field und the sample. If the repulsion force $F_r$ is equal in amount and opposite in direction to the gravitational force, $F_r = m\, g$, the sample is levitated. $m$ denotes the mass of the sample and $g$ the gravitational acceleration. Electromagnetic levitation can be used to levitate metallic and even semi-conducting samples. However, electromagnetic levitation of semiconductors requires either doping with a suitable element to increase the electrical conductivity or preheating the pure semiconductor to a temperature of about 1000 K by a laser or by a graphite susceptor within the levitation coil so that the intrinsic conduction is sufficiently increased to electronically couple the sample to the alternating external field. A characteristic feature of electromagnetic levitation is that both levitation and heating of the sample are always occurring simultaneously. This offers the advantage that no extra source of heating is required to melt the material, but it is associated with the disadvantage that levitation and heating can be controlled independently only in a very limited range.

According to Rony [45] the mean force on an electrically conductive non-ferromagnetic sample is determined by

$$F_{em} = -\frac{4\pi r}{3} \cdot \frac{B \cdot \nabla B}{2\mu_o} \cdot G(q) \tag{2.3}$$

Here, $r$ denotes the radius of the sphere-like sample, $\mu_o$ the permeability of vacuum. The function $G(q)$ is calculated as

$$G(q) = \frac{3}{4}\left(1 - \frac{3\sinh(2q) - \sin(2q)}{2q\cosh(2q) - \cos(2q)}\right) \tag{2.4}$$

q is the ratio of the sample radius $r$ and the skin depth $\delta$

$$q = \frac{r}{\delta} \quad \text{with} \quad \delta = \sqrt{\frac{2}{\mu\omega\sigma}} \tag{2.5}$$

$\omega$, $\sigma$ and $\mu$ are the angular frequency of the electrical current, the electrical conductivity and the magnetic permeability of the sample, respectively. According to equation (2.3) the levitation force, $F_L = F_{em}$, scales with the gradient of the magnetic field. To optimize levitation, it is therefore crucial to design properly the geometry of the levitation coil and optimize the function $G(q)$. The efficiency of electromagnetic levitation is adjusted by the parameters of the frequency of the alternating electromagnetic field, the sample size and the electrical conductivity of the sample. For a vanishing conductivity ($q \rightarrow 0$) $G(q)$ becomes zero and levitation is not possible. For ($q \rightarrow \infty$) $G(q)$ is approaching saturation.

To levitate a sample of masse $m$ the gravitational force $F_g$ has to be compensated by the electromagnetic levitation force $F_L$

$$F_L = -F_g; \quad F_g = m \cdot g = \frac{4\pi r^3}{3} \rho \cdot g \tag{2.6}$$

where $\rho$ denotes the mass density of the material. The z-component of the force follows as

$$\frac{\partial B^2}{\partial z} = \frac{2\mu_o g}{G(q)} \cdot \rho \tag{2.7}$$

For a given magnetic field and sample size the levitation force is determined by the skin depth $\delta$ and the mass density $\rho$. The mean power absorption $P$ is calculated according to Roney as

$$P = \frac{B^2}{2\mu_o} \cdot \omega \cdot \frac{4\pi r^3}{3} \cdot H(q) \qquad (2.8)$$

with

$$H(q) = \frac{9}{4q^2} \cdot \left( q \cdot \frac{\sinh(2q) - \sin(2q)}{\cosh(2q) - \cos(2q)} - 1 \right) \qquad (2.9)$$

$H(q)$ is the efficiency of the power absorption. For vanishing electrical conductivity no power is absorbed by the sample. On the other hand for an ideal conductor no ohmic losses occur so that $H(q)$ converges to zero.

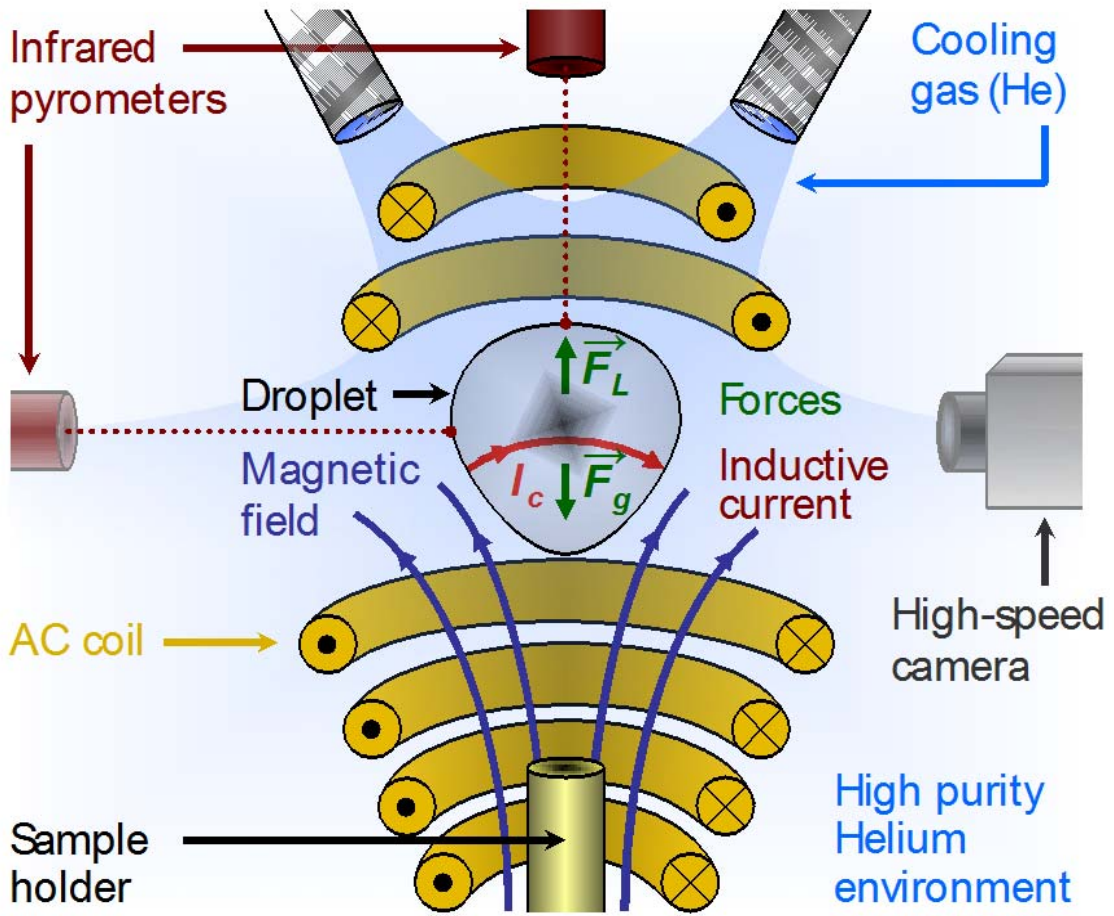

**Figure 1.** Schematics of electromagnetic levitation; the droplet is placed in a levitation coil consisting of four water cooled windings and two counter windings at the top to stabilize the position of the levitated drop. Two infrared pyrometers measure contactless the temperature of the droplet from the top and from a radial side. A high-speed camera is used to observe the rapid propagation of the solidification front. Cooling gas from the top is necessary to remove the heat from the sample to be cooled and undercooled. The arrows give the geometry of the magnetic field inside the coil. $F_L$ and $F_g$ denote the levitation force and the gravitational force while $I_c$ is the current induced in the droplet by the alternating electromagnetic field. Cooling of the sample requires convective

cooling of a gas of high thermal conductivity like helium. (Reproduced with permission from Acta Materialia 59 (2011) 4665–4677. Copyright 2011 Elsevier)

The temperature control of electromagnetically levitated samples requires a separate action of $P$ and $F_L$. The essential difference between $P$ and $F_L$ is that the functions $G(q)$ and $H(q)$ have a different characteristics with respect to the frequency of the alternating electromagnetic field. Further on, $F_L$ depends on the product $(B\bullet\nabla)B$, while $P$ is proportional to $B^2$ (cf. eqs. 2.3 and 2.8). Hence, temperature control is possible within a limited range by choosing a proper frequency of the alternating field and by a movement of the sample along the symmetry axis of a conically shaped coil. In the lower regions of the coil the windings are tighter, thus the magnetic field and power absorption are greater than in the upper region of the coil with lower field strength. By increasing the power the sample is lifted up into regions of larger field gradients but smaller magnetic field strength and cools down [3].

**2.1.2 Electromagnetic levitation in reduced gravity**

The application of electromagnetic levitation on Earth is limited by several restrictions. The strong electromagnetic fields needed to compensate the gravitational force cause strong stirring effects in the liquid and, hence, disturb mass and heat transport that influence solidification. The strong electromagnetic levitation fields exert a magnetic pressure on the liquid sample that leads to deviations from a sphere-like geometry, which however is needed for measurements of surface tension and mass density. These limitations are overcome if electromagnetic levitation technique is applied in reduced gravity. In the environment of space the forces to compensate disturbing accelerations are of some orders of magnitude smaller compared with experiments on ground. A special instrument called TEMPUS[1] has been designed to provide means of containerless processing in space [46].

A schematic view of the TEMPUS concept is shown in Figure 2. Positioning and heating is separated in TEMPUS by placing the sample into the superposition of a quadrupole and a dipole field of two independent coil systems. Both coil systems are powered independently by two rf generators at different radio frequencies. This two-coil concept has led to a drastic increase of the heating efficiency of levitated drops from about 1% on Earth to about 30%using the TEMPUS concept a mandatory precondition for Space experiments with their limited energy resources. The coil system is integrated in a UHV chamber. The recipient can be backfilled with high purity Ar, He and/or He-3.5%$H_2$ processing gas. Solidification of the undercooled melt can be externally triggered by touching the undercooled sample with a nucleation trigger needle, which is an integral part of the sample holder. The samples are transferred into the coil system from sample storage within ceramic cups or refractory metal cages.

TEMPUS is equipped with pyrometers and video cameras. The sample is observed from two orthogonal views. From the top a pyrometer measures the temperature with a frequency of 1kHz. A video camera is included in the optical path for sample observation

---

[1] TEMPUS is a German acronym for <u>T</u>iegelfreies <u>E</u>lektro-<u>M</u>agnetisches <u>P</u>rozessieren <u>U</u>nter <u>S</u>chwerelosigkeit

with a maximum frame rate of 400 Hz. From the side two different instruments can be installed, either a pyrometer specialized for measurements of the crystal growth velocity at rates up to 1 kHz (RAD), which is combined with a video camera with frame rates up to 400 Hz, or a high-resolution video camera (RMK) with special optics. The resolution is $10^{-4}$ for a 8mm sample as required for measurements of the thermal expansion. TEMPUS was successfully flown by NASA Spacelab missions IML2 (International Microgravity Laboratory 1994) and MSL1/MSL1R (Materials Science Laboratory 1997).

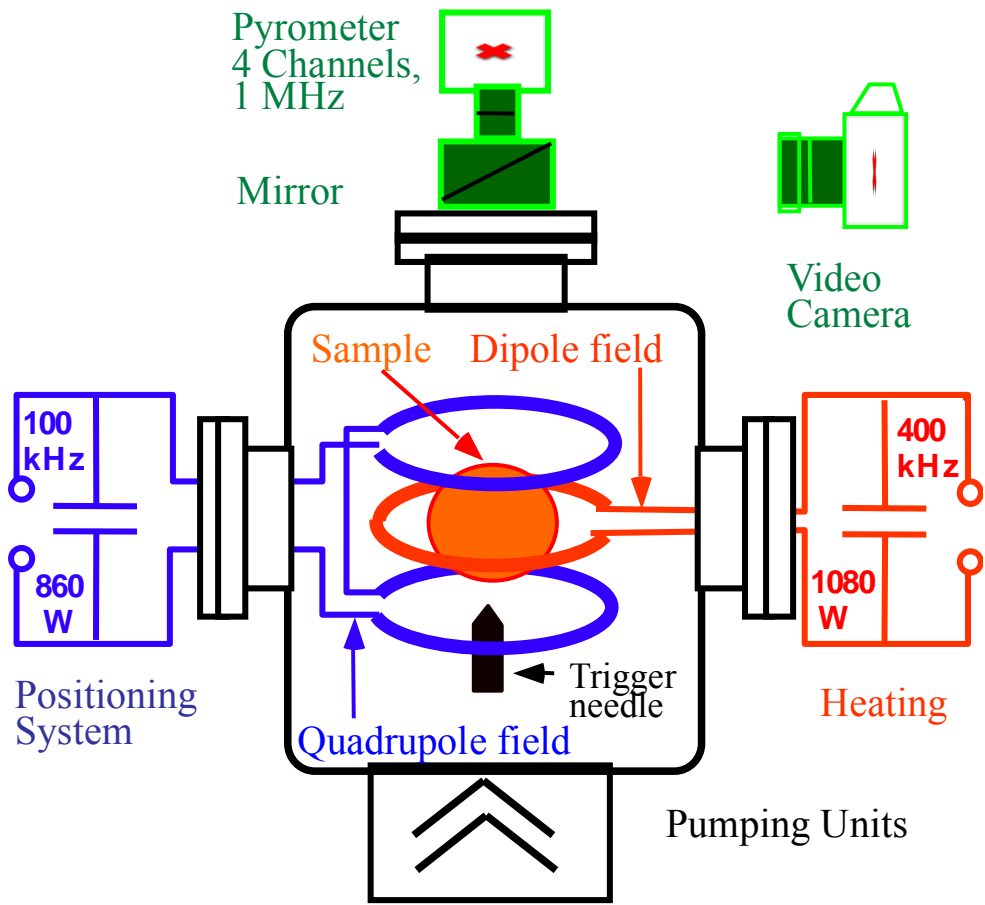

**Figure 2:** Schematic view of the TEMPUS facility. All subsystems are shown with the exception of the radial temperature detector.

In a common effort by DLR Space Agency and ESA an Electro-Magnetic Levitator (EML) has been developed. It is accommodated on board the International Space Station (ISS). Comparing with TEMPUS some important improvements are realized. The first one concerns the coil design. While TEMPUS used two different coils, the EML facility will make use of a new concept [47] such that only one coil system carries two different high frequency alternating currents. The first one operates at a frequency of 135 – 155 kHz and serves as positioning system whereas the second one operates at a frequency of 365-390 kHz and provides efficient heating. At maximum power the positioning force in radial direction is about 80% and in axial direction about 120% of the force of the two-coil system of TEMPUS. The heating efficiency of EML is by a factor of 1.6 higher and the maximum heating power in the sample is about 30% higher compared with TEMPUS. The EML facility is equipped with axial temperature measurement and video observation of the sample. The temperature of the sample is measured by an one color pyrometer in axial direction in the range between 573 K and 2373 K at an integration time of 5 ms and an accuracy better than 0.1 K at temperatures greater than 873 K and less than 3.0 mK in the temperature range 573 K – 673 K. The measurement rate amounts to 100 Hz. An axial digital video camera allows to observing the sample during processing. The maximum resolution is 1280 x 1024 pixels and the frame rate ranges between 15 Hz and 200 Hz depending on the pixel density. In addition to the pyrometer and video system working in axial direction a high-speed camera is used to observe the propagation of a rapidly moving solidification front. The maximum measuring frequency is 30 kHz at a pixel density of 256 x 256 pixels. Also, thermal radiation monitoring from radial view is possible. At present, several international teams are preparing experiments of different classes to be performed in the EML facility on board the International Space Station [12].

**2.1.3 Electrostatic levitation**

Electromagnetic levitation requires sample material that is electrically conductive. Therefore, the application of electromagnetic levitation is restricted to metals and (doped) semiconductors. The advantage of electrostatic levitation is that levitation and heating is decoupled and the samples can be processed under ultra high vacuum conditions provided the vapor pressure of the processed material is small. However, there is a problem with the stability of the sample position. According to the theorem of Samuel Earnshow, there is no potential valley for a stable position of a charged sphere within a static electrostatic field [48]. For electrostatic levitation a sophisticated sample position observation system and a fast responding high voltage source is needed.

Electrostatic levitation is based on the Coulomb forces acting on an electrically charged sample in a quasi-static electrical field [49]. A sample with a surface charge $q$ and a mass $m$ is levitated against gravity within a static electrostatic field $\vec{E}$ as

$$\vec{E} = -\frac{mg}{q} \cdot \vec{e}_z \quad (2.10)$$

$\vec{e}_z$ is the unit vector in z direction, i.e. parallel to the electrostatic field. A stable position of the sample is based on a local potential minimum at $\vec{r}_o$ for all directions in space.

$$\frac{\partial^2}{\partial x^2}\Phi(\vec{r}_o) + \frac{\partial^2}{\partial y^2}\Phi(\vec{r}_o) + \frac{\partial^2}{\partial z^2}\Phi(\vec{r}_o) = \Delta\Phi(\vec{r}_o) > 0 \quad (2.11)$$

The Maxwell-equation for Gauss's law affords

$$\Delta\Phi = -\frac{\rho}{\varepsilon_o} \quad (2.12)$$

Under vacuum conditions, $\Delta\Phi = 0$. Hence, a potential minimum does not exist and a stable sample position under stationary conditions is not possible, i.e. electrostatic levitation requires a dynamic sample position and electrostatic field control. This became possible just since the nineties of last century when high voltage amplifiers were developed, which can be controlled with high slew rates of changing the voltage $U$, $dU/dt > 400V/\mu s$.

Figure 3 shows schematically the active sample positioning system. An electrically charged sample is levitated between two horizontal electrodes within a widened positioning laser beam filling the whole space between the electrodes. The sample shadow is detected by a two-dimensional photo-sensitive detector that gives information on the vertical and horizontal position of the sample. A real time computer control algorithm developed by Tilo Meister [50] reads this information and adjusts instantaneously the voltage of the amplifier. In order to control the sample position in all three dimensional directions two positioning laser perpendicular to each other and an assembly of six electrodes are used. Two central electrodes arranged as a plate capacitor are surrounded by four electrodes in plane, which are cross-linked with the positioning lasers to push the sample in the central position. The forces acting in z-direction, $F(z)$, are the gravitational force, the force due to the electrical field, and the force between the sample and the grounded center electrodes. With the method of image charges the force of a charged sphere between the electrodes can be determined by

$$F(z) = \frac{q^2}{4\pi\varepsilon_o}\left(\sum_{n=1}^{\infty}\frac{1}{(2d_z n - 2z)^2} - \sum_{n=0}^{\infty}\frac{1}{(2d_z n - 2z)^2}\right) \qquad (2.13)$$

with the position of the sample $z$, the distance of the electrodes $d_z$, the charge $q$ of the sample, the vacuum permittivity $\varepsilon_o$, and the number of reflections $n$. Neglecting multiple reflections, $F(z)$ is approximated as

$$F(z) \approx -\frac{q}{4\pi\varepsilon_o}\left(\frac{1}{(2z)^2} - \frac{1}{(2d_z - 2z)^2}\right) \qquad (2.14)$$

In the middle of the electrodes the forces of the image charge acting on the sample are compensating each other. The equation of motion for the z direction is given by

$$m\ddot{z} = -mg - q\frac{U_z}{d_z} - \frac{q^2}{4\pi\varepsilon_o}\left(\frac{1}{(2z)^2} - \frac{1}{(2d_z - 2z)^2}\right) \qquad (2.15)$$

The field in x-direction and y-direction are assumed to be between two parallel electrodes

$$m\ddot{x} = -2q\frac{1}{\kappa}\cdot\frac{U_x}{d_x} \quad \text{with} \quad \kappa = \sqrt{\left(\frac{2z}{d_x}\right)^2 + 1} \qquad (2.16)$$

$d_x$ is a geometrical factor regarding the distance of the sample and the lateral electrodes.

For conducting an experiment using the electrostatic levitator the sample in diameter of about 2-4 mm is placed at the lower electrode, which is grounded. The high voltage power supply is switched on and immediately an electrostatic field between upper and lower electrode in z-direction is built up. At the same time the sample is charged. Since the upper electrode is on negative potential the surface of the sample is loaded with positive charge $q^i$ that is calculated as [51]

$$q^i = 4\pi\varepsilon_o L \frac{U_z^i}{d_z} r^2 \qquad (2.17)$$

with r the radius of the sample and L = 1.645 a geometrical factor. The image charge of the bottom electrode dominates the initial levitation voltage. The force acting on a sample while lifting is given by

$$F_z^i = \frac{4}{3}\pi r^3 \rho g - q\frac{U_z^i}{d_z} - \frac{q^2}{4\pi\varepsilon_o} \cdot \frac{1}{(2r)^2} = 0 \qquad (2.18)$$

Combining eq. (2.17) and eq. (2.18) yields the initial voltage for levitation

$$U_z^i = \pm d_z \sqrt{\frac{4\rho g r}{3L(4-L)\varepsilon_o}} \qquad (2.19)$$

The charge of the sample in the beginning of the experiment is then

$$q^i = \mp 8\pi \sqrt{\frac{L\varepsilon\rho g r^3}{3(4-L)}} \qquad (2.20)$$

The voltage $U_z^o$ needed to keep the sample in the middle of the electrodes is calculated as

$$U_z^o = \frac{4-L}{4} U_z^i \qquad (2.21)$$

The initial voltage is larger than the voltage that is needed to levitate the sample in the middle of the horizontal electrodes. For a constant initial voltage the time is approximated which elapses until the sample hits the electrode. This time is used to estimate the minimum sampling rate required for positioning. For a silicon sample in diameter of 2mm the sampling rate is $2 \cdot 10^{-3}$ s.

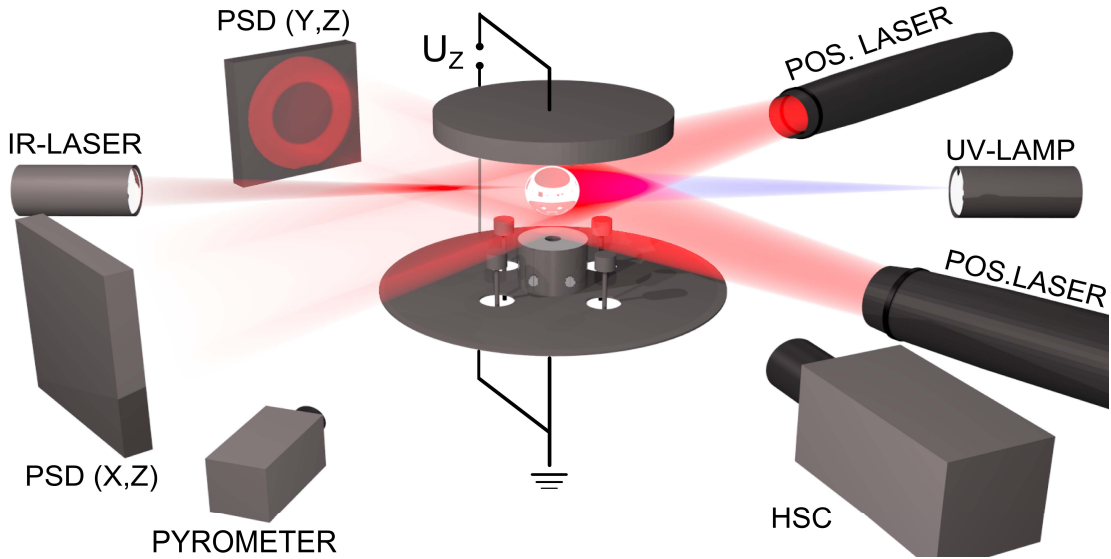

**Figure 3:** Sketch of the fully automated electrostatic levitator, consisting of a symmetrical designed system of electrodes with an upper and a lower electrode for levitation (z-axis) and four electrodes in the horizontal plane (xy-plane) for positioning. Due to the absence of position stabilizing minima in an arrangement of fixed electric charges the samples position has to be adjusted in real-time which is ensured by the shadow of the sample cast during orthogonal illumination with a pair of He/Ne-lasers onto oppositely mounted photosensitive detectors (PSD). During levitation the sample is heated and melted with an infrared laser (P = 75W, λ = 808nm) and the temperature is measured contactless with a pyrometer (Impac IGA 120-TV, 1000 Hz, accuracy ±5K). (Reprinted with permission from the PhD Thesis Work Raphael Kobold, *Crystal growth in undercooled melts of glass forming Zr-based alloys*, Ruhr-University Bochum (2016)).

Electrostatic levitation offers the advantage that positioning and heating are decoupled in contrast to electromagnetic levitation. Heating is realized in electrostatic levitation by an infrared laser. Increasing the temperature of the sample leads to an evaporation of surface atoms, which is useful for undercooling experiments since the evaporation cleans the surface and thereby reduces or even eliminates heterogeneous nucleation motes at the surface of the sample. On the other hand, the sample surface looses surface charge by evaporation. Therefore, the voltage has to be increased to keep the sample levitated. To facilitate recharging of the sample during levitation a focused ultra violet light source with a high energy of several eV (wavelength λ=115-350nm) is used. In addition to this procedure, the sample is also recharged at elevated temperatures by thermionic emission of electrons [52].

The electrostatic levitator is very suitable to study nucleation undercooling with special emphasis to homogeneous nucleation. To observe homogeneous nucleation very large undercoolings have to be realized, since the onset of homogeneous nucleation gives the physical limit for maximum undercoolability of a melt. To realize such conditions heterogeneous nucleation has to be eliminated. Electrostatic levitation under ultra-high-vacuum is ideally suited for such experimental studies since heterogeneous nucleation on container walls is completely avoided and heterogeneous nucleation on surface motes is reduced or even eliminated due to self cleaning of the surface by evaporation at elevated temperature.

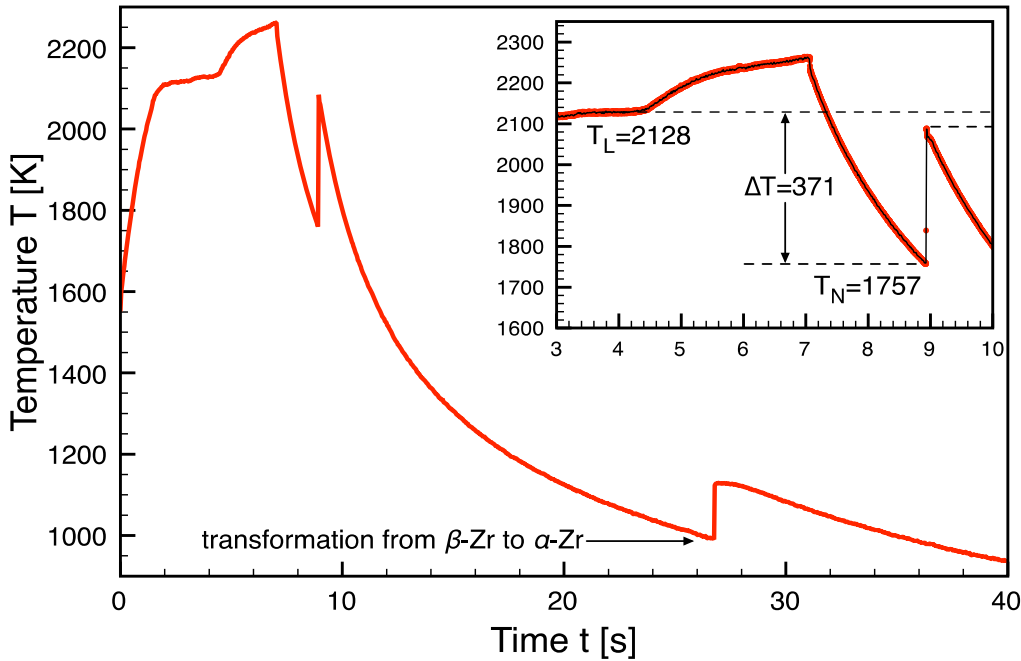

**Figure 4:** Temperature-time profile measured on a Zr drop in an electrostatic levitator. The sample melts at $T_L$=2128 K. During undercooling nucleation sets in at $T_N$=1757 K. Subsequently, rapid crystal growth of β-Zr solid phase (bcc) leads to a steep rise of temperature during recalescence. The second recalescence event at 980 K is attributed to a transformation of solid β-Zr to solid α−Zr

phase (hcp). The inset shows the determination of the undercooling ΔT = $T_L$ - $T_N$. (Reprinted with permission from PhD Thesis Work Stefan Klein, *Nucleation in undercooled melts of pure zirconium and zirconium-based alloys*, Ruhr-University Bochum (2010))

Nucleation undercooling studies on pure Zr are presented to demonstrate how physically different nucleation processes are experimentally investigated. Figure 4 shows a temperature-time-profile measured on pure Zr sample in the electrostatic levitator (cf. inset). The solid sample is heated up to its melting temperature, $T_L$. In case of a pure metal as Zr the sample melts congruently at $T_L$ i.e. no temperature change during melting. The small step in the melting plateau is due to the change in spectral emissivity when the solid transforms to the liquid. After complete melting the liquid sample is heated to a temperature well above $T_L$ before cooling. During subsequent cooling the liquid sample undercools well below $T_L$. When spontaneous nucleation sets in at an undercooling $\Delta T = T_L - T_N$ ($T_N$: nucleation temperature) the nucleated crystal rapidly grows due to a large thermodynamic driving force generated at such deep undercoolings. The rapid release of the heat of crystallization leads to a steep rise of temperature during recalescence. From such temperature-time profiles $\Delta T$ is easily inferred since the nucleation temperature $T_N$ is well defined by the onset of recalescence. After the entire sample has solidified the next heating and cooling cycle is started

The solidification of an undercooled metallic melt is a two-staged process. During recalescence a fraction of the sample, $f_R$, solidifies under non-equilibrium condition. The remaining melt, $f_{pr}=1- f_R$, solidifies under near-equilibrium conditions during post-recalescence period. $f_R$ increases with the degree of undercooling and becomes unity, $f_R=1$ if $\Delta T = \Delta T_{hyp}$. The hypercooling limit, $\Delta T_{hyp}$, is reached if the heat of fusion $\Delta H_f$ is just sufficient to heat the sample with its specific heat $C_p$ up to $T_L$. In case of quasiadiabatic conditions, i.e., if the amount of heat transferred to the environment is negligible compared to the heat produced during recalescence, the hypercooling limit is given by $\Delta T_{hyp}=\Delta H_f/C_p$. In case of pure Zr the hypercooling limit is estimated as $\Delta T_{hyp}$=359 K with $\Delta H_f$=14652J/mol and $C_p$ =40.8 J mol/K. With increasing undercooling, $\Delta T > \Delta T_{hyp}$, the post-recalescence plateau vanishes and $T_L$ will not be reached during recalescence. In this experiment an undercooling of $\Delta T$=371 K is measured, which is larger than $\Delta T_{hyp}$.

## 2.2 Colloids

### 2.2.1 Preparation of charge stabilized colloidal suspensions

A colloidal experiment aiming at a comparison to atomic systems starts with the choice of a suitable model type interaction. In case condensation is studied, colloids with entropic interactions are the preferred choice [53], as they show an attractive component in the potential of mean force, which mimics Lennard-Jones potentials [36]. In the case of metals, charged colloidal spheres are ideally suited. In monodisperse, highly charged colloidal systems the effective pair interaction energy *V(r)* can be written as:

$$V(r) = \frac{(Z_{eff}e)^2}{4\pi\varepsilon_0\varepsilon r}\left(\frac{\exp(-\kappa a)}{1+\kappa a}\right)\frac{\exp(-\kappa r)}{r} \qquad (2.22)$$

with the screening parameter $\kappa$

$$\kappa = \sqrt{\frac{e^2}{\varepsilon_0 \varepsilon_r k_B T}(n_P Z_{eff} + 2000 N_A c)} \tag{2.23}$$

Here, $Z_{eff}$ is an effective charge, $e$ is the elementary charge, $a$ is the particle radius, $\varepsilon_o \varepsilon_r$ is the dielectric permittivity of the suspension, $k_B T$ is the thermal energy, $n_P$ is the particle number density, $N_A$ is Avogadro's constant and $c$ is the molar concentration of electrolyte. In Eqns. (2.22) and (2.23) one recognizes the familiar form of a Yukawa-potential, which, however, contains a term for finite size correction. With a grain of salt, the particles act as big, highly charged atomic nuclei, while the role of the screening electrons is taken by the highly diffusive, classical counter-ions in the electric double layer. This combination favorably mimics the repulsive part of the metal-metal interaction.

With a few exceptions [54, 55], the interaction does not strongly depend on temperature, as the individual dependencies neatly compensate. However, such systems allow for a convenient variation of the pair interaction energy through changes of particle radius $a$, particle number density $n$ and electrolyte concentration $c$. Therefore, utilization of a continuous conditioning technique is a prerequisite for each measurement [56]. It is used to guarantee a reproducible adjustment of interaction parameters by varying and controlling the particle number density and the counter ion concentration [57]. The suspension is pumped peristaltically through a closed tubing system connecting an ion exchange chamber to deionize the sample, a reservoir under inert gas atmosphere to add further suspension or water, a conductivity cell in order to determine the salt concentration and measuring cells for scattering or microscopy experiments [58].

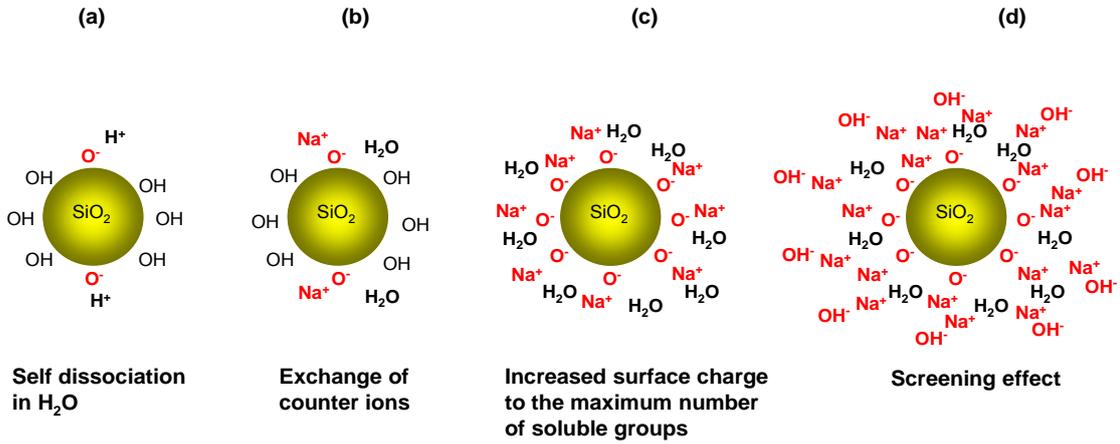

**Figure 5**: Control of silanol surface groups on silica particles and their interaction as function of added NaOH with increasing NaOH concentration from left to right. (Reprinted with permission from PhD Thesis Work Ina Klassen, *Charged colloids as model systems for metals*, Ruhr-University Bochum (2009)).

Variation of the charge is also easily performed, although its theoretical description needs some careful considerations concerning the hierarchy of different types of charge. The final effective charge is first of all determined by the colloidal material used. Silica systems can for instance be synthesized employing a modified Stöber synthesis [59]. This method allows for the synthesis of spherical silica particles with small size polydispersity (about 5%) in a size ranging from 5 to 2000 nm. In the present studies we have utilized two different batches of home made Silica particles with diameters of e.g. 77 nm (Si77) and 84 nm (Si84). They are small enough for Brownian motion avoiding sedimentation in the gravitational field, but just large enough for accurate scattering studies. Furthermore the silica colloids show a high scattering contrast for X-rays even at low volume fractions.

The synthesis of the silica colloids include a hydrolysis reaction

$$Si(OR)_4 + H_2O \xrightarrow{OH^-} (OR)_3Si(OH) + ROH \qquad (2.24)$$

to produce the single-hydrolyzed TEOS monomer *(OR)$_3$Si(OH)*. Subsequently, this intermediate reaction product condenses to form silica

$$(OR)_3SI(OH) + H_2O \longrightarrow SiO_2 \downarrow + 3ROH \qquad (2.25)$$

The synthesized *SiO$_2$* particles carry weakly acidic silanol groups (Si – OH) on the surface, which partly dissociate in a de-ionized water environment leaving spheres with a negative surface charge. Surface groups of particles with strong acidic end groups are fully dissociated with their maximum possible surface charge. The dissociation of silanol groups by which silica surfaces acquire a charge in contact with water is described by SiOH→SiO$^-$ + H$^+$.

The dissociation equilibrium of silanol groups is strongly depend on the pH of the surrounding media. To influence the surface charge of silica particles and thus the electrostatic interaction of the particles sodium hydroxide (NaOH) is added to the suspension. The addition of NaOH controls the degree of dissociation and the surface charge density [60]. By adding sodium hydroxide the counter ion species first changes from H$^+$ to Na$^+$ followed by a charging of the silica particles due the reaction *Si-OH + NaOH → Si-O$^-$ + Na$^+$ + H$_2$O*. The particles charge up until their maximum surface charge is achieved and the bare or structural charge $Z$ equals the group number N. Therefore, the surface charge density $\sigma_a$ can be calculated from the added amount of base as [61]

$$\sigma_a = \frac{10^{-3}}{3} N_A \cdot e \cdot a \frac{c_{NaOH}}{\Phi} \qquad (2.26)$$

Here, $\Phi=(4/3)\pi a^3$ is the volume fraction of the silica particles. The equivalence point of the titration corresponds to the point of maximum interaction. Further increase of the NaOH concentration, the so-called excess concentration, causes a screening of the particle surface charge and the interaction decreases. This mechanism is shown schematically in Figure 5. It should further be noted, that at least large particles alternatively could be studied by colloidal probe AFM. There, force-distance curves are directly measured and interpreted in terms of a charge regulation model [62].

Use of the bare charge Z in the pair potential is fully sufficient only on the level of isolated particle pairs. Notably, the superposition approximation implicitly used in Eqn. (2.22) has been verified in both tweezing experiments [63] and numerical simulations [64]. However, as soon, as the double layer of a third particle overlaps with those of the pair, many body corrections have to be considered which weaken the electrostatic repulsion. Two effects are most apparent. The first is related to a change in the double layer structure. It has been termed 'counter-ion condensation' and is theoretically captured using a renormalized charge Z* [65, 66]. The predicted scaling of Z* with the particle radius R [65] has been demonstrated in several experiments on isolated particles [67, 68] and its saturation with increasing Z has been shown in neutron scattering experiments on mixed ionic-non-ionic micelles [69]. The renormalized charge is further accessible as an effective charge from electro-kinetic experiments in general [70, 71, 72] and conductivity experiments in particular [73, 74, 75]. Such experiments probe single particle properties under conditions of overlapping double layers.

Second, the charge of the additional particles introduces an additional screening-like effect termed 'macro-ion shielding' [76]. For a triplet of particles, these effects render the interaction anisotropic. Their description requires many body corrections to be introduced [77, 78] which get ever more complex as the number of neighbors is increased further. Experimentally, it was shown that in the limit of a finite density of particles, a spherically symmetric potential of mean force is recovered, which appears to be effectively truncated at the nearest neighbor distance [76]. It may further lead to anisotropic elastic properties in single crystals [79, 80]. At large $n_P$, however, the truncation effects by macro-ion shielding become negligible, as classical screening by counter-ions decreases the potential already over short distances [81]. As an effect of interest in the present studies, the renormalized charge is further reduced to an effective charge $Z_{eff}$. Also this charge is experimentally accessible. Observations by video-microscopy [82], measurements of the osmotic pressure [83], the structure [84, 85, 86, 87] and of the shear modulus of polycrystalline solids [88, 89] have been successfully employed for that purpose. As it turns out, a consistent description of the phase behavior of charged particles becomes feasible, if the effective charge from elasticity is used as input for a prediction of the melting line of the suspension utilizing the ('universal') freezing line obtained in simulations [90, 91]. This works remarkably well for one-component suspensions [92] even in the case of variable charges [23], but fails for binary mixtures, where the corresponding phase diagram predictions are still missing [93].

**2.2.2 Laser light scattering**

As introduced above, the colloid specific length scales allow for a convenient but powerful approach via optical methods like microscopy or light scattering. Both yield complementary information from real and reciprocal space. Using optical methods the structural properties, particle dynamics and the phase transition kinetics of colloidal model systems as well as the elastic properties of colloidal solids (and so the particle interaction) can be investigated with high precision allowing a full characterization of the colloidal model system.

A specific mechanical property of colloidal solids is their softness. Due to the low particle number density $n$ of some $10^{18}$-$10^{19}$ m$^{-3}$ (atomic systems $n \approx 10^{29}$ m$^{-3}$) the elastic moduli G are in the range of only a few Pa (atomic systems G $\approx 10^{10}$-$10^{11}$ Pa). The fragility of such samples is a major constraint of investigations and it is mandatory to have a high precision light scattering set-up, which enables simultaneous investigations of different properties without disturbing the sample. A multi-purpose light scattering device has been developed by T. Palberg and H. J. Schöpe [94, 95, 96] that combines quasi simultaneous static and dynamic light scattering (SLS and DLS) probing the structure and morphology of colloidal solids, respectively their dynamics, with torsion resonance spectroscopy (TRS) to determine the shear modulus of polycrystalline samples and by this way the particle interaction.

We show a sketch of the apparatus in Figure 6. The set up comprises of a NdYag-laser working at $\lambda_0$ = 532 nm, an index match bath for the sample cell, as well as of separate sending and detection optics for SLS and DLS, all mounted on a vibrating-free optical table. The sample cell made of quartz glass is in the center of the index match bath also made of quartz glass. Index matching is necessary to avoid parasitic reflections and correct for the refractive widening of the scattered beam.

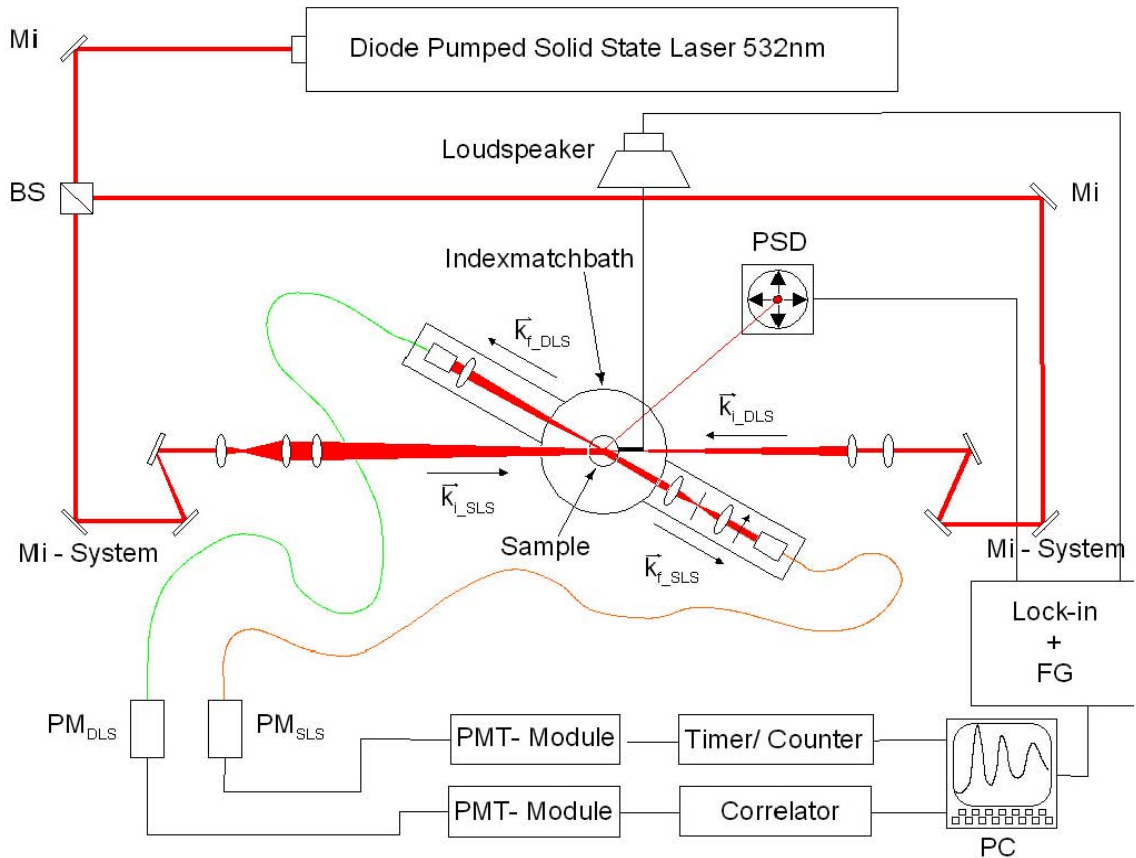

**Figure 6:** Sketch of the multi-purpose device for simultaneous measurements of static, dynamic and elastic properties of colloidal systems. The basic construction of the multipurpose device in particular the double-arm goniometer for the simultaneous measurements of static, dynamic and elastic properties was reported in [94]. The figure includes following abbreviations: mirror (Mi), beam splitter (BS), photomultiplier (PM) and position-sensitive detector (PSD). The PC screen shows a resonance spectrum. (Reprinted with permission from PhD Thesis Work Ina Klassen, *Charged colloids as model systems for metals*, Ruhr-University Bochum (2009)).

To realize a counter-propagating illumination and inversion symmetric detection from the very same particle ensemble, the laser light is split by a beam splitter and injected into two exactly opposing sending optics. For SLS illumination a broad, parallel beam is required, to assure good powder averaging while for DLS the Gaussian beam waist (diameter about 100μm) must pass exactly through the cell center. The detectors are placed on a double-arm goniometer with exactly opposing arms. In Figure 6, the laser beam for DLS enters from the right and is detected using the fiber optics of the upper arm. The intensity autocorrelation function is calculated from the output signal of the photomultiplier. The beam for SLS and torsion resonance detection enters from the left and is detected using the SLS optimized fiber optics of the lower arm. The position sensitive detector used for TRS can be freely positioned and is adjusted to collect light of a single Bragg reflection on the Scherrer cone of a powder sample. For TRS the sample is set into gentle torsional oscillations about its vertical axis by means of a mechanically coupled loud speaker driven by a frequency generator. The driving frequency and the response measured by the position-sensitive detector are fed into a lock-in amplifier to record the resonance spectrum. From this, the shear modulus is derived and, assuming the effective pair interaction described in Eqn. 2.22, an effective charge, $Z_{eff}$ is calculated. Due to the fact that silica suspensions are not optically matched, light scattering is restricted to lower particle concentration where multiple scattering effects can be neglected. At high particle concentration meaningful measurement by light scattering are quite difficult [97, 98, 99, 100, 101] and X-Ray scattering is an alternative method obtaining data from the reciprocal space

**2.2.2 Ultra small Angle X-ray scattering**

With increasing particle number density the colloidal suspension becomes impervious to light in the visible spectrum that makes measurements by optical light quite difficult. Change of the wavelength from the optical range to the ultraviolet range is of no help since colloidal suspensions absorb ultraviolet light. The use of X-rays opens up an extended field of scattering experiments. Since the wavelength of X-ray is much smaller than the interparticle distance in colloidal suspensions X-ray scattering experiments have to focus on Ultra Small Angle X-ray Scattering (USAXS). In addition, USAXS makes use of the penetration of X-rays through materials, either in solid or liquid state. Typical scattering angles for USAXS measurements range around 1° leading to a much larger scattering vector range compared to light scattering experiments.

The beamline BW4 at HASYLAB in Hamburg is an X-ray wiggler (N=19 periods, K=13.2) beamline with instrumentation to perform Ultra Small Angle X-ray Scattering especially on soft matter systems with a distance between sample and detector of 13.5 m. Details are given elsewhere [102]. The experimental arrangements are shown in Figure 7.

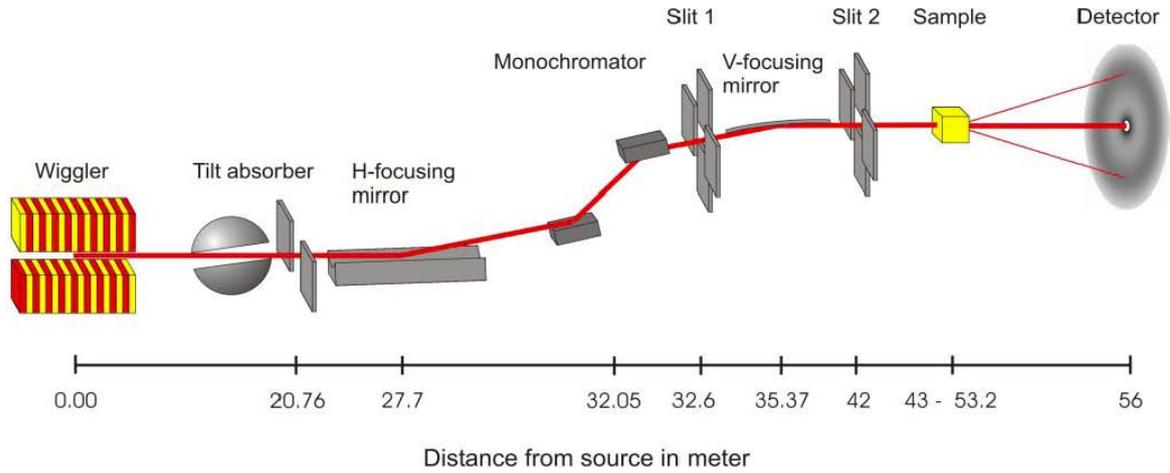

**Figure 7:** The beam-line BW4 at HASYLAB (DESY) [102]. The beam-line starts at the location of the Wiggler BW4 and has a total length of 56 m. The tilt absorber is the only element in the beam-line capable to absorb the total power of the white wiggler beam. The horizontally focusing optical element (H-focusing mirror) is a water-cooled platinum coated silicon mirror with fixed cylindrical shape. The BW4 has a fixed-exit double-crystal-monochromator with Si(111) crystals. The vertically focusing optical element (V-focusing mirror) is a plane mirror installed in a mirror-bending device to realize a curved mirror surface. The guard slit system (Slit 1 and 2) consists each of four silicon slit blades driven by a piezo motor with an accuracy of 0.1μm. The blades have a special geometry and surface treatment to reduce scattering from the slits. (Reprinted with permission from PhD Thesis Work Ina Klassen, *Charged colloids as model systems for metals*, Ruhr-University Bochum (2009)).

Typical for transmission USAXS is the use of a beam stop obscuring the direct beam. A photodiode in the center of the beam stop is used to monitor the primary beam intensity passed through the sample. The resolution is limited by the minimum size of the beam stop resulting in the maximum detectable correlation distance of $d_{max}$ = 650 nm. The beam stop prevents the detector from being damaged by the direct beam. For monitoring the intensity of the incident primary beam an ionization chamber is provided. Two types of detectors are applied for detecting the scattered intensity. The first is a commercial CCD-detector (MARCCD165) with an active diameter of 165 mm and a resolution of 2048 x 2048 (pixel size 79.1 x 79.1 $\mu m^2$) which operates with a read-out time of 3.5 sec. In addition, a PILATUS100K detector with an active area of 84 x 34 mm and a resolution of 487 x 195 (pixel size 172 x 172 $\mu m^2$) with a read-out time of 5 x $10^{-3}$ sec was utilized. PILATUS detectors have several advantages in comparison to CDD and image plate detectors including a high frame rate up to 100 Hz, no read-out noise, excellent signal-to-noise ratio and highly detected quantum efficiency. These properties ensure a good image quality even at short exposure times.

A key parameter characterizing SAXS measurements is the resolution in terms of the minimum accessible scattering angle $\theta_{min}$. By Bragg's equation it is related with the maximum correlation distance $d_{max}$ accessible in real space:

$$d_{max} = \frac{\lambda}{2\sin\theta_{min}} \approx \frac{\lambda}{2\theta_{min}} \qquad (2.27)$$

where λ denotes the used wavelength of x-rays. The minimum scattering angle of the BW4 setup is on the order of 0.01°. At particle distances at 360 nm, the largest possible sample-to-detector distance has to be chosen and is calibrated with a collagen sample to be 13.3m. With these parameters, the accessible range of scattering vectors is 0.8 µm$^{-1}$<q<320µm$^{-1}$ corresponding to real space length scales 3 nm < d < 1 µm.

In pre-evaluation, the scattering pattern is divided by the actual incident flux measured by the ionization chamber and by the exposure time t. Further evaluation of the diffraction patterns includes a background and transmission correction of the 2D scattering intensities. For the background correction, the scattering intensity $I_{bg}(x;y)$ of the solvent and of the window material without the colloidal suspension is measured and has to be subtracted from the scattering profiles measured with the colloidal sample. The transmission correction considers the sample thickness and the fact that X-rays are partially absorbed when they pass through matter. As a result, the total transmitted intensity $I_t$ measured after passing the sample is only a fraction $I_t/I_0$ of the incident intensity $I_0$. In combination with the absorption law, following absorption and background correction has been applied:

$$I_{corr}(x,y) = I_s(x,y)\exp\left(\frac{\mu l}{\cos 2\theta}\right) - I_{bg}(x,y) \qquad (2.28)$$

$I_{bg}(x,y)$ and $I_s(x,y)$ where $I_s(x;y)$ and $I_{bg}(x;y)$ are the measured 2-dim intensities with and without the colloidal sample. Here µ is the linear absorption coefficient (as a function of the X-ray wavelength and the chemical composition of the sample) and $l$ is the sample thickness. The intensity is increasingly dampened with increasing scattering angle. Simplification is possible in the case of small scattering angles (2θ →0 and thus cos 2θ → 1) obtaining the linear absorption factor exp(-µl). The experimental determination of the absorption factor is based on measurements of photon counts before (N$^1$) and behind (N$^2$) the sample. In this case, $\exp(-\mu l) \approx \left(N_s^2/N_s^1\right)/\left(N_{bg}^2/N_{bg}^1\right)$ is approximately valid leading to the following expression for the transmission and background correction

$$I_{corr}(x,y) \approx I_s(x,y)\left(\frac{N_s^2 N_{bg}^1}{N_s^1/N_{bg}^2}\right) - I_{bg}(x,y) \qquad (2.29)$$

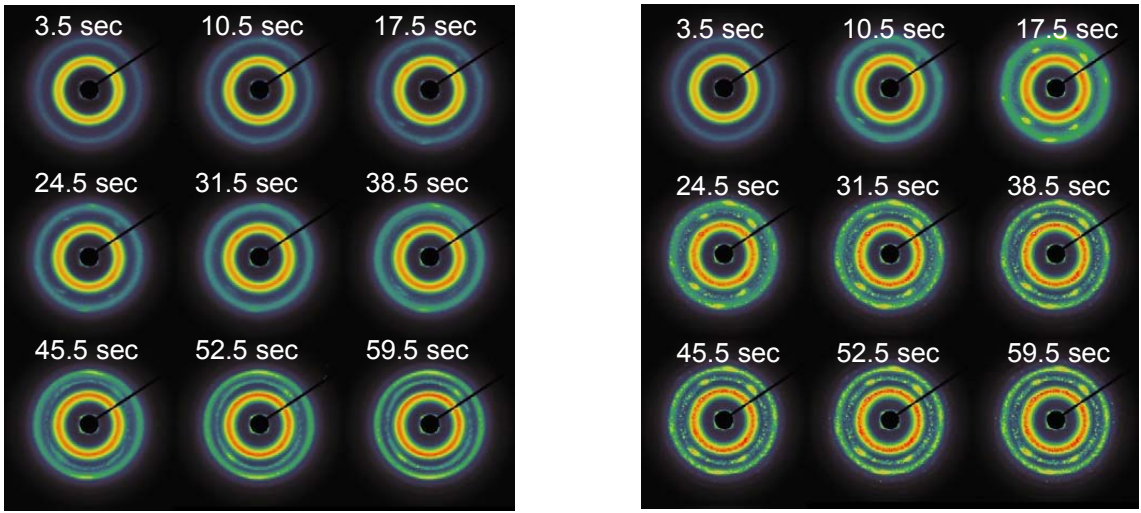

**Figure 8:** Sequence of nine diffraction patterns of colloidal silica suspensions with particle diameter 84 nm and particle number density n=113μm$^{-3}$ at strongest particle interaction adjusted via the sodium hydroxide concentration $c_{NaOH}$=1.06·10$^{-3}$ mol/l (left) and weaker interaction at smaller concentration $c_{NaOH}$=0.4·10$^{-3}$ mol/l (right). Homogeneous nucleation dominates in the suspension of maximum interaction. The four-fold diffraction spots on the right indicate that heterogeneous nucleation of oriented wall crystals becomes dominant if the particle interaction is weakened [95]. (Reprinted with permission from PhD Thesis Work Ina Klassen, *Charged colloids as model systems for metals*, Ruhr-University Bochum (2009)).

Time resolved diffraction patterns of the shear molten and re-crystallizing colloidal silica suspension measured for several interaction strengths are shown in Figure 8. Two examples of these 2-dim sequences are exhibited for a silica suspension (Si 84) at a particle number density n = 113 μm$^{-3}$. The exposure time and the read-out time of the detector are both 3.5 sec in this case.

The left Figure 8 represents the system at maximum interaction ($c_{NaOH}$= 1.4·10$^{-3}$ mol/l) where homogeneous nucleation dominates the crystallization process. The first diffraction pattern shows the meta-stable melt state. After the third diffraction pattern, the Debye-Scherrer rings become narrower and additional rings appear indicating the transition to a polycrystalline bcc state. In addition, from the beginning of crystallization a weakly developed four-fold point pattern is visible which can be ascribed to an oriented wall based single crystal originated from heterogeneous nucleation [30, 103]. Heterogeneous nucleation is more pronounced in the state of lower interaction strength. This effect is shown in the right part of Figure 8 where a small amount of added NaOH ($c_{NaOH}$= 0.4 10$^{-3}$ mol/l) is present. Here, the appearance of heterogeneous nucleation has a larger influence on the scattering profiles. Both diffraction patterns show one of the main advantages of USAXS measurements of colloidal systems, namely the discrimination of the scattering information resulting from polycrystalline material or oriented wall crystals.

Wette et al. analyzed the competitive behavior between heterogeneous and homogeneous nucleation near a flat wall [95]. Accordingly, heterogeneous nucleation at the container walls is delayed in comparison with homogeneous nucleation within the bulk fluid. The structure factor *S(q)* was obtained from the intensity of the diffracted beam *I(q)* by considering the particle form factor *P(q)* as

$$I(q) \propto I_o \cdot n \cdot P(q) \cdot S(q) \qquad (2.30)$$

The particle form factor *P(q)* is determined by proper calibration USAXS measurements. $I_o$ is the intensity of the incident beam, and *n* the particle number density of the suspension used.

# 3. Experimental results and discussions – a representative selection

## 3.1 Metals and alloys

### 3.1.1 Short range order in the undercooled liquid state of pure metals and alloys

The structure factor of undercooled melts of pure metals as Ni, Fe, Zr [8] and Co [104] was measured in the stable regime and below $T_L$ at undercooling up to 140 K for Fe and even to 290 K for Ni and Zr. This corresponds to high relative undercooling ($\Delta T = T_L - T$) especially in the case of Ni (17%). Such a large temperature range should allow one to evidence significantly the structural changes occurring in the liquid *S(Q)*. As an example Figure 9 shows the measured structure factor of pure Ni at a temperature of T=1453 K (solid circles) that is 276 K below the melting temperature. A shoulder on the right-hand side of the second oscillation is observed. This shoulder is characteristic of a particular short-range order (SRO). Such a feature has been identified as a signature of icosahedral short range-order (ISRO) from the theoretical work of Sachdev and Nelson [105] and was evidenced in quasicrystal forming alloy melts [106]. This feature becomes more pronounced as *T* is lowered and undercooling is increased. Inter-atomic distances and coordination numbers are quantitative parameters of SRO obtained from the corresponding pair correlation functions *g(R)*. The nearest neighbor distance, *R*1, remains essentially unchanged while the second neighbor distance, *R*2, slightly decreases with decreasing *T*. The nearest neighbor coordination number, $Z \approx 12$ is obtained by the area under the first maximum. It was determined by integration over the whole area between first and second minima limiting this first oscillation. The Z values ($Z \approx 12$) obtained are characteristic of several types of densely packed SRO, e.g., icosahedral, fcc, and hcp.

For analysis of the measured structure factor an approach was used to simulate the structure factor S(Q) for large *Q* values. It is based on the assumption that aggregates of specific structure are present which do not interact with each other. It implies only three free fit parameters to discriminate between the different structures of aggregates: bcc, fcc, hcp, icosahedral, and dodecahedral. The simulation method is outlined in detail in [106]. The parameters of the simulation are the shortest mean distance, $\langle r_o \rangle$, of atoms contained in a cluster, its mean thermal variation, $\delta r$, that determines the Debye-Waller factor, $\exp(-\langle \delta r_i^2 \rangle)$, and the concentration, *X*, of cluster atoms in the liquid. The mean thermal variations of the other intra-cluster distances, $\langle r_i \rangle$, are estimated by

$$\langle \delta r_i^2 \rangle = \frac{\langle \delta r_o^2 \rangle \langle r_i^2 \rangle}{\langle r_o^2 \rangle}$$

The parameters are adjusted such that a good fit of the experimentally determined $S(Q)$ is obtained especially at high $Q$ values. The large $Q$ part of $S(Q)$ is determined by the SRO only as far as the contribution of larger distances is damped by thermal motions.

The results of simulations for liquid Ni in the undercooled regime at $T = 1435$ K for differently structured aggregates are presented in Figure 9. For a SRO of bcc- and hcp-like structure it is not possible to achieve a reasonable fit of the measured $S(Q)$. The best fit is obtained for icosahedral-like SRO. In particular, the assumption of ISRO can better reproduce the large oscillation at about 55 Å and the presence of the characteristic shoulder as an asymmetry on the right-hand side of this oscillation.

We have been dealing up to here with isolated 13 atom aggregates, but we may wonder about the nature of SRO at larger distances. A simulation based on a larger cluster, the dodecahedron, was performed. The dodecahedron is an aggregate of higher order consisting of 33 atoms, which can be constructed from the icosahedron by placing atoms densely on all its 20 triangular faces. When comparing the results obtained for the two clusters with icosahedral symmetry it is obvious that the assumption of dodecahedral aggregates leads even to a better description of the measured $S(Q)$. This may indicate that a SRO order consisting of larger poly-tetrahedral aggregates (such as dodecahedra) dominates in the undercooled liquid.

The fivefold symmetry of the ISRO is not compatible with the translational symmetry of crystals as they solidify from pure metals. The dissimilarity between SRO in the liquid and the solid may lead to a large interfacial energy, and thus to a large activation energy to form supercritical nuclei. In his early work Frank has used this argument to explain the large undercooling [107], which have been experimentally observed by application of the volume separation of heterogeneous nucleation sites [108]. On the other hand, if SRO order is similar in liquid and solid the activation energy for nucleation should be small. As a consequence, undercooling will be limited and smaller than for pure metals.

Quasicrystal forming alloys represent a class of solids, which lack long-range translational symmetry of a crystal but show long range orientational order. The maximum undercooling attainable for quasicrystal forming alloys of different degrees of fivefold symmetry has been investigated by levitation experiments. It was found that the higher the degree of fivefold symmetry, the smaller was the maximum undercooling of the respective alloys [109]. From this, it is concluded that such alloys will show fivefold SRO in the state of an undercooled melt. Figure 10 shows, as an example, the partial Bhatia-Thornton structure factor $S_{NN}$ of the quasicrystal forming $Al_{13}(Co,Fe)_4$ alloy in the liquid undercooled state measured by elastic neutron scattering. The analysis of the experimental data leads to the conclusion, that, in fact, this alloy shows a topological SRO of fivefold symmetry quite similar to the pure metals [110]. Consequently, the maximum relative undercooling, $\Delta T_R = T_L - T_n / T_L$ ($T_L$ liquidus and $T_n$ nucleation temperature) of these alloys are by a factor of two smaller than the typical relative undercooling of metals

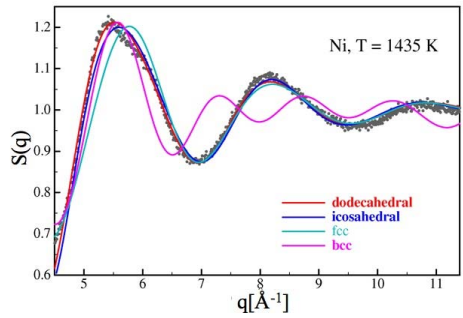 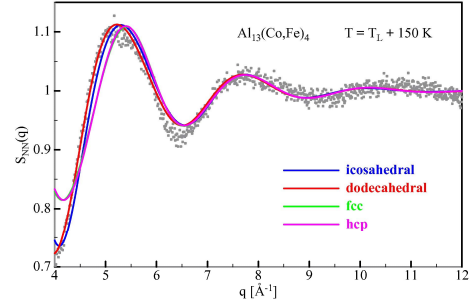

**Figure 9:** Structure factor S(Q) measured by elastic neutron scattering on levitation undercooled Ni sample at a temperature T=1435 K, that corresponds to an undercooling of 286 K. Circles represent experimental data, lines give the results of simulations assuming bcc, fcc, icosahedral and dodecahedral short-range order, respectively. (Reproduced from T. Schenk, D. Holland-Moritz, V. Simonet, R. Bellisent, D. M. Herlach, *Icosahedral short-range order in deeply undercooled metallic melts*, Physical Review Letters **89** (2002) 075507-1-075507-4, with the permission of APS Publishing, copyright 2002 APS).

**Figure 10:** Structure factor S(Q) measured on quasicrystal forming alloy $Al_{13}(Fe,Co)_4$ in liquid state. The symbols represent the experimental data whereas the lines give the results of simulations assuming bcc, fcc, icosahedral and dodecahedral SRO, respectively. (Reproduced from T. Schenk, V. Simonet, D. Holland-Moritz, R. Bellissent, T. Hansen, P. Convert, D.M. Herlach, *Temperature dependence of the chemical and topological short-range order in undercooled and stable Al-Fe-Co liquids*, Europhysics Letters **65** (2004) 34-40, with permission of EPS Publishing, Copyright 2004 EPS).

The situation again changes, if glass-forming alloys are studied. As a side-remark, this topic has also been extensively studied in colloidal glass formers [111]. In both system classes, glass formation requires the avoidance of crystal nucleation. For both hard sphere suspensions and alloys, it was argued in the literature that the preference of ISRO in the undercooled liquid favors the glass-forming ability of such systems [42, 112]. In fact a series of Zr-based amorphous metals shows the primary formation of a fivefold symmetric icosahedral i-phase upon crystallization [113, 114, 115]. However, concerning icosahedral short-range order in the liquid state of Zr-based glass-forming alloys, there is some controversy in the literature. X-ray scattering experiments using high-energy synchrotron radiation on $Zr_{80}Pt_{20}$ alloys undercooled by electrostatic levitation reveal a structural factor S(Q) that shows a pre-peak and a second peak with a shoulder [116].

Figure 11 compares the structure factor of pure Zr undercooled by 298 K with the structure factor S(Q) of intermetallic compound $PdZr_2$ undercooled by 250 K [117]. The melt of pure Zr shows a clear preference of icosahedral short-range order (Figure 11) quite similar as for pure Ni. By contrast to the pure metal, the measured structure factor of the intermetallic compound $PdZr_2$ suggest rather a mixture of differently structured aggregates to be present in the state of the undercooled melt (Figure 12) with an average coordination number Z = 13.8. Concerning the $PdZr_2$ compound, one may conclude that the liquid consists of a variety of local structural aggregates of different short-range order. Since the coordination number obtained for $PdZr_2$ is larger than the coordination number obtained for Zr some of the aggregates must be either more densely packed than the icosahedral or close-packed crystal-type clusters or more highly coordinated. Ab initio MD simulations are in agreement with this assumption, suggesting that multiple differently structured aggregates are present in undercooled liquids of Zr-Cu [118] and Ni-Zr [119] glass-forming alloys. Moreover, the results are in agreement with a RMC analysis of X-ray scattering data on a $Zr_{75.5}Pd_{24.5}$ alloy that was constrained based on *ab initio* molecular dynamics results [120] which found a very broad distribution of cluster types with only a modest abundance of complete icosahedral clusters. Depending on the chemical and topological structures of the aggregates relative to those of the energetically accessible crystal phases, this may lead to difficulties in crystal nucleation, therefore favoring glass formation. Both samples have been largely undercooled. At such a deep undercooling, the asymmetry of the second peak is most pronounced that has been proposed to be a strong indication of icosahedral short-range order [121]. However, it is noteworthy that the asymmetry of the second peak is also clearly visible of the structure factor of the inter-metallic compound despite the fact that in the undercooled melt of $PdZr_2$ the icosahedral order may be present but it does not dominate. This result is opposite to investigations of crystallization of amorphous $PdZr_2$ alloys. Murty et al. have suggested the concept of the preference of icosahedral short range order in the liquid to analyse nucleation studies on amorphous Pd-Zr alloys with composition close to the intermetallic $PdZr_2$ intermetallic compound [122].

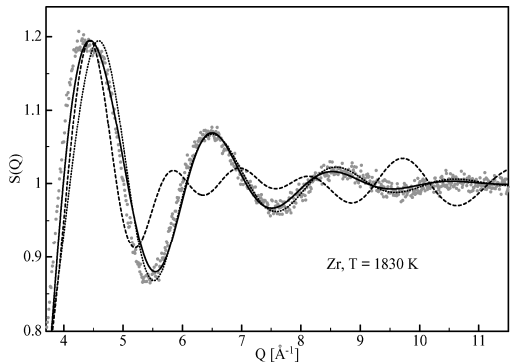 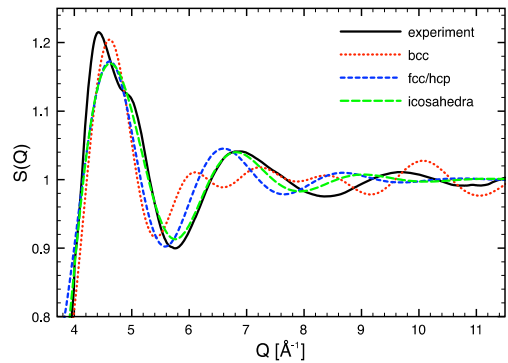

**Figure 11:** Structure factor of pure Zr measured by neutron scattering at undercooling ΔT=298K (black dots), computed for bcc (dashed), fcc/hcp (dotted) and icosahedral (solid line) ordering. A preference of icosahedral short range order is obvious by the best agreement of computed and measured structure factor. (Reprinted with permission from PhD Thesis Work Stefan Klein, *Nucleation in undercooled melts of pure zirconium and zirconium-based alloys*, Ruhr-University Bochum (2010)).

**Figure 12:** Structure factor of intermetallic compound $PdZr_2$ measured by X-ray scattering using synchrotron radiation at undercooling of ΔT=248K (black), computed for bcc (red), fcc/hcp (blue) and icosahedral (green) ordering. No preference of any short range order is visible, as apparent by the disagreement of all computed structure factors with the experiment. (Reprinted with permission from PhD Thesis Work Stefan Klein, *Nucleation in undercooled melts of pure zirconium and zirconium-based alloys*, Ruhr-University Bochum (2010)).

### 3.1.2 Phase selection by crystal nucleation

An electromagnetic levitation chamber is used to combine it with external diagnostic means for e.g. neutron scattering and X-ray scattering by synchrotron radiation [6]. The primary nucleation of a meta-stable bcc phase in Ni-V alloys at large undercooling was directly evidenced by in situ energy dispersive X-ray diffraction on levitation processed undercooled melt using high intensity synchrotron radiation at the European Synchrotron Radiation Facility [11]. This becomes possible since a full diffraction spectrum is recorded in a very short time interval less than 0.5 seconds.

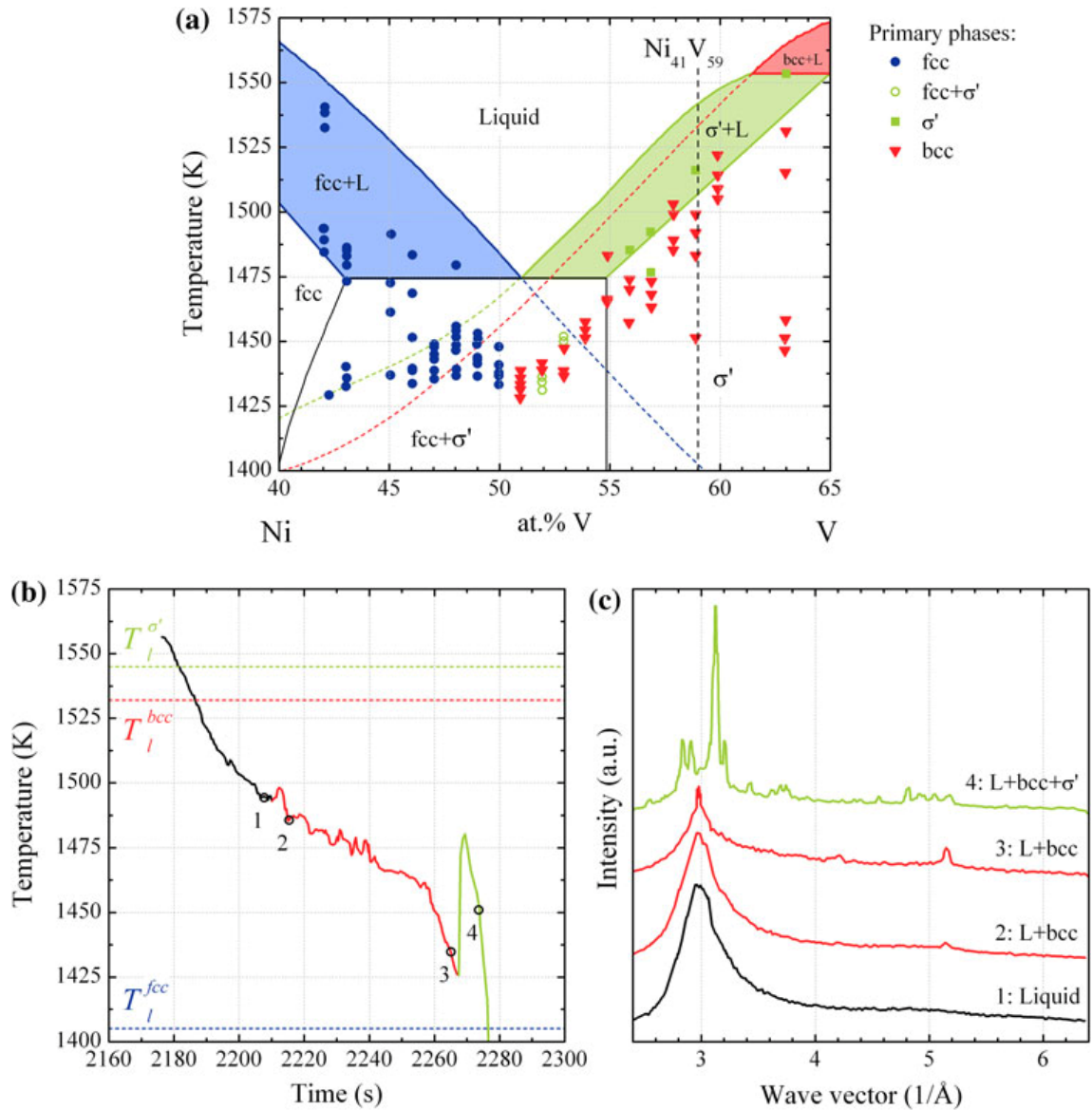

**Figure 13:** (a) nucleation map of Ni–V system after. The symbols mark the primarily solidified phases as a function of composition and nucleation temperature. Metastable extensions of the equilibrium liquidus lines are shown by dotted lines. (b) Temperature–time profile and (c) corresponding X-ray diffraction patterns recorded during solidification of a levitated $Ni_{41}V_{59}$ alloy. Numbers 1–4 mark the time at which the patterns are recorded. Related liquidus temperatures of stable σ' phase and metastable bcc and stable fcc phases are shown. A part of the undercooled melt (pattern 1) primarily nucleates into the metastable bcc phase (pattern 2 and 3) followed up by the nucleation of the inter-metallic σ' phase (pattern 4). (Reprinted with permission from O. Shuleshova, W. Löser, D. Holland-Moritz, D.M. Herlach, J. Eckert, *Solidification and melting of high temperature materials: in situ observations by synchrotron radiation*, J. Mater. Sci. 47 (2012) 4497-4513, Copyright 2012 Springer).

By systematic investigations as a function of concentration and undercooling using this diagnostic means complete meta-stable phase diagrams have been experimentally constructed as exemplary shown in Figure 13 for Ni-V alloy system [7]. This alloy crystallizes in three different phases depending on the concentration. With increasing vanadium fraction the equilibrium solidification changes from primary α-Ni (fcc) to an inter-metallic phase σ' (tetragonal) and eventually to β-V (bcc).

Upon undercooling the three phases compete with each other. The metastable bcc β-V

phase has been primarily nucleated beyond a critical undercooling within the ranges of the stable α-Ni and σ' phase. Diffraction pattern 1 taken from a melt undercooled by 50 K below the liquidus temperature $T_l$ (σ ) shows diffuse maxima characteristics of the liquid. Diffraction pattern 2 reveals that at $\Delta T = 60$ K part of the undercooled melt crystallized into the metastable bcc phase. The bcc phase grows continuously upon further cooling (diffraction pattern 3) until the second reaction sets in at $\Delta T = 120$ K, where reflections of the equilibrium σ' phase emerge. At the same time the bcc phase completely disappears either by re-melting or via a liquid–solid reaction. For lower vanadium concentrations only primary nucleation of the equilibrium fcc phase was observed, even though for some compositions the undercooling was well below $T_{lm}$ (β) —the metastable extension of the liquidus temperature of the β-V phase. However, after several seconds, the bcc phase was formed during the second crystallization event. The nucleation and growth of the metastable bcc phase did not influence the intensity of the fcc phase reflections, thus one may conclude it grows from the undercooled melt independently. Again the meta-stable phase decomposes after crystallization of the remaining liquid into a mixture of the equilibrium phases, σ' and fcc. Although the observed formation of the metastable bcc phase is thermodynamically viable, its nucleation before the σ' phase, which in general possesses a lower solid– liquid interfacial energy, can only be explained by the catalytic effect of the oxides, triggering the heterogeneous nucleation of the bcc phase at the droplet surface [11].

These investigations unambiguously demonstrate that in situ observations of phase selection processes are needed to develop a complete understanding of the nucleation of different crystallographic phases stable or meta-stable during non-equilibrium solidification of deeply undercooled melts while post mortem analyses of as solidified samples are not sufficient to explore the entire crystallization history. Primarily formed crystalline phases can transform to the stable crystalline phase during the post-recalescence period by re-melting processes or even during cooling by solid-solid phase transformations. Quenching the levitated drop immediately after recalescence allows to freezing in the primarily crystallized meta-stable phase as reported from similar experimental investigations of phase selection of steel alloys and refractory metals [123].

### 3.1.3. Crystal nucleation in the undercooled melt of pure Zr and binary Zr-based compound

The short-range order in the undercooled melt is often considered as precursor of crystal nucleation. In particular, icosahedral short-range order should hinder crystal nucleation but should favor the solidification of quasicrystalline phases. This means if thermodynamics allows for the formation of crystals of translational symmetry only the Zr melt with the preference of icosahedral short range order should undercool considerably provided heterogeneous nucleation of high catalytic potency can be avoided. On the other hand, the presence of aggregates of cubic symmetry in undercooled $NiZr_2$ melt may facilitate crystal nucleation of crystalline phases of similar symmetry and therefore the maximum attainable undercooling should be decreased in this case. To investigate the undercooling behavior of both different systems both electromagnetic and electrostatic levitation are utilized. In particular, electrostatic levitation is attractive to reach an undercooling, which is limited by intrinsic homogeneous nucleation since experiments are conducted by containerless proc-essing under ultra-high-vacuum conditions on small droplets. The nucleation undercooling distribution functions of pure Zr and NiZr$_2$ compound have been investigated by undercooling and solidifying each sample at least 100 times comparatively by both levitation techniques. The experimental results are analyzed by using the Skripov model for the statistics of nucleation undercooling [124]. Nucleation events are considered to be independent events. The Poisson distribution expresses the discrete probability distribution $\omega$ of a number of nucleation events $m$ occurring in a fixed period of time if these events happen with a known average rate

$$\omega(m,\tau) = \frac{(\lambda \tau)^m}{m!} \exp(-\lambda \tau) \tag{3.1}$$

Under the assumption that one nucleation event is sufficient to initialize solidification it holds:

$$\int_0^{t_N} J_{ss} \cdot V \, dt_N = 1 \tag{3.2}$$

with $J_{ss}$ the steady state nucleation rate, $t_N$ the time that elapses until a nucleus initializes solidification at the nucleation temperature $T_N$. Under non-isothermal conditions with cooling rate (dT/dt ≠ 0) the probability distribution ω(T+δT) is extracted from the Poisson distribution. The probability of nucleus inception at a time interval τ, τ+δτ is given by λ(τ)δτ. The probability of the lack of this event at $\tau$ is determined by

$$P(0,\tau) = \lambda(\tau) d\tau \cdot \exp\left(-\int_0^\tau \lambda(\tau) d\tau\right) \tag{3.3}$$

The probability for one nucleation event occurring within the temperature interval T+δT is computed as

$$P(1, T+\delta T) = \partial T \frac{J_{ss} V}{\dot{T}} \cdot \exp\left(-\int_{T_L}^{T} \frac{J_{ss} V}{\dot{T}} dT\right) \tag{3.4}$$

with

$$\omega(T) = \frac{P(1, T+\delta T)}{\partial T} = \frac{J_{ss} V}{\dot{T}} \cdot \exp\left(-\int_{T_L}^{T} \frac{J_{ss} V}{\dot{T}} dT\right) \tag{3.5}$$

Neglecting a temperature dependence of the specific heat difference between solid and liquid, the Gibbs free energy between both phases is described by the linear approximation $\Delta G_V = \Delta S_f \Delta T / V_m$ with $\Delta S_f$ the entropy of fusion and $V_m$ the molar volume. Neglecting a temperature dependence of the viscosity in the liquid below the liquidus temperature far above the glass transition temperature the steady state nucleation rate $J_{ss}$ is written as

$$J_{ss} = K_V \exp\left(-\frac{\Delta G^*}{k_B T}\right) = K_V \exp\left(-\frac{CT^2}{\Delta T}\right) \tag{3.6}$$

with the activation energy

$$\Delta G^* = \frac{16\pi}{3} \frac{\gamma^3}{\Delta G^2} f(\vartheta) \quad \text{and} \quad C = \frac{16\pi}{3} \frac{\Delta S_f \alpha^3 f(\vartheta)}{k_B N_A}$$

$N_A$ is Avogradro's number, $\gamma$ the interfacial energy, $f(\theta)$ the catalytic potency factor of heterogeneous nucleation and $\Delta G$ the difference of Gibbs free energy of solid and liquid. $\alpha$ denotes the dimensionless interfacial energy, $\alpha = \gamma/\Delta S_f$.

Eventually, we arrive at the cumulative distribution function

$$F(T) = 1 - \exp\left( \frac{V K_V}{\dot{T} \dfrac{d\left(\dfrac{\Delta G^*}{k_B T}\right)}{dT}} \cdot \exp\left(\frac{CT^2}{\Delta T^2}\right) \right) \tag{3.7}$$

According to this equation a plot of

$$\ln(-\ln(1 - F(T)))$$

versus

$$T^2 / \Delta T^2$$

yields a linear relation with slope $-C$ and intercept b

$$C = \frac{\Delta T^3}{2(T\Delta T + T^2)} \cdot \frac{d\left(\dfrac{\Delta G^*}{k_B T}\right)}{dT}; \quad b = \ln\left(\frac{V K_V}{\dot{T} \dfrac{d\left(\dfrac{\Delta G^*}{k_B T}\right)}{dT}}\right) \tag{3.8}$$

From this formalism, the activation energy $\Delta G^*$ to form a critical nucleus and the pre-factor $K_V$ of the nucleation rate can be determined. With these parameters the probability distribution function is computed and compared with the experimental results.

The procedure of this analysis is described in detail elsewhere [125]. This analysis provides figures of the most important parameters of nucleation, that is the activation energy to form a nucleus of critical size, $\Delta G^*$, and the pre-factor of the nucleation rate, $K_V$, that gives a measure of potential nucleation seeds and nuclei formed per volume unit and time unit. In case of homogeneous nucleation and collision limited attachment kinetics of atoms to the solid nucleus Turnbull estimated the pre-factor $K_V = 10^{39}$ m$^{-3}$s$^{-1}$ in case of homogeneous nucleation [126]. More recently, Dantzig and Rappaz propose an even larger value of $K_V = 10^{42}$ m$^{-3}$s$^{-1}$ [127]. The activation energy $\Delta G^*$ is inferred from the average undercooling and the pre-factor from the half width of the computed distribution function

The results of both experimental and theoretical investigations for pure Zr [125] are shown in Figure 14, and those of NiZr$_2$ in Figure 15, respectively. For comparison, also results are shown which have been achieved on pure Zr using electromagnetic processing in reduced gravity during the NASA Spacelab mission MSL1 in 1997 [128]. They are not

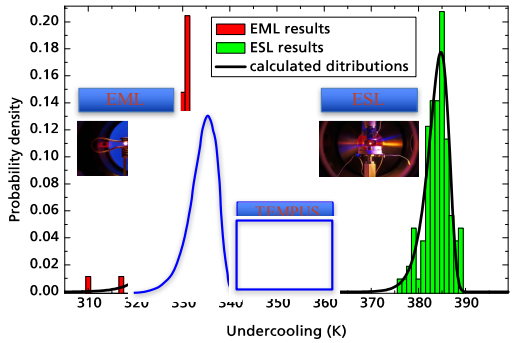 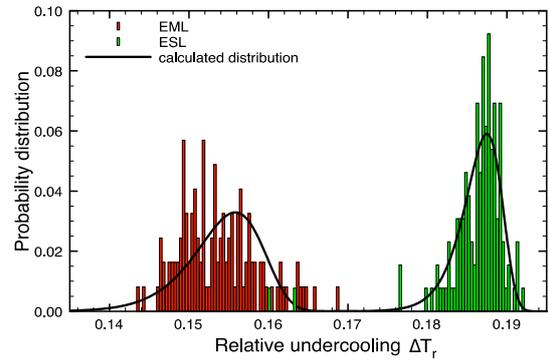

**Figure 14**: Probability distribution functions of the relative undercooling measured in approximately 100 solidification cycles on pure Zr in the electromagnetic on Earth (left), electromagnetic levitator in Space (middle), and electrostatic (right) levitator. The solid lines give the results of computations according to an analysis of the nucleation statists. (Reprinted with permission from PhD Thesis Work Stefan Klein, *Nucleation in undercooled melts of pure zirconium and zirconium-based alloys*, Ruhr-University Bochum (2010)).

**Figure 15**: Probability distribution functions of the relative undercoolings $\Delta T_r = \Delta T/T_L$ measured on $NiZr_2$ with EML(left) and ESL (right). The solid lines give the probability distribution function as computed according to the Skripov model. (Reprinted with permission from PhD Thesis Work Stefan Klein, *Nucleation in undercooled melts of pure zirconium and zirconium-based alloys*, Ruhr-University Bochum (2010)).

The maximum undercoolings measured for pure Zr in the electrostatic levitation yield a relative undercooling $\Delta T_r$, of about 18% of the respective melting temperature $T_L$, $\Delta T_r = 0.18\ T_L/\Delta T$ with $\Delta T$ the undercooling measured in the individual solidification cycle. This is comparable with the maximum relative undercooling obtained by Turnbull for a great spectrum of pure metallic elements applying the droplet dispersion technique [108]. Figure 15 shows results of nucleation undercooling statistics obtained for the intermetallic compound $NiZr_2$ in EML and ESL experiments. The results obtained for the $NiZr_2$ compound are similar as those obtained for pure Zr. In case of $NiZr_2$ the experiments in the ESL yield systematically larger values of undercooling and the half width of the distribution function is broader. One may argue that the high purity environmental conditions are responsible that the undercooling measured in ESL are higher than those measured in EML. It is noted that the experiment conditions are throughout comparable as being present in equivalent experiments on $PdZr_2$ but in the latter experiments a larger undercooling is achieved by electromagnetic levitation instead of the electrostatic levitation as it the case of NiZr$_2$ compound [129]. It is not yet understood that the results of the comparative EML and ESL experiments on NiZr$_2$ and PdZr$_2$ are just opposite to each other

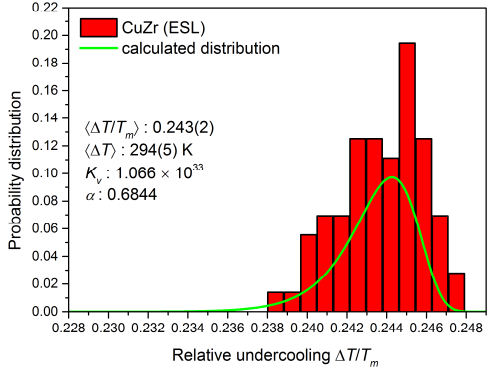

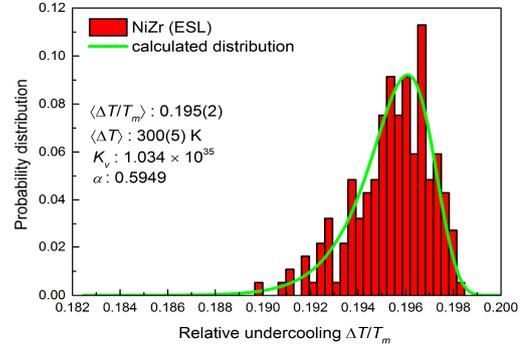

**Figure 16:** Probability distribution functions of the relative undercoolings $\Delta T_r = \Delta T/T_L$ measured on CuZr with ESL and ESL. The solid lines give the probability distribution function as computed according to the Skripov model [130]. (Reprinted with permission from PhD Thesis Work Raphael Kobold, *Crystal growth in undercooled melts of glass forming Zr-based alloys*, Ruhr-University Bochum (2016)).

**Figure 17:** Probability distribution function of the relative undercoolings $\Delta T_r = \Delta T/T_L$ measured on NiZr with ESL. The solid line gives the probability distribution function as computed according to the Skripov model [130]. (Reprinted with permission from PhD Thesis Work Raphael Kobold, *Crystal growth in undercooled melts of glass forming Zr-based alloys*, Ruhr-University Bochum (2016)).

**Table 1:** Parameters collected from statistical analysis of nucleation undercooling measured for more than 100 cycles for pure Zr, and inter-metallic compounds NiZr$_2$, NiZr, CuZr. These are the average undercooling, $\Delta T_{avg}$, the relative undercooling $\Delta T_r$, the activation energy to form critical nuclei, $\Delta G^*$, and the pre-factor of the nucleation rate $K_V$. Results of Zr and NiZr$_2$ are taken from ref. [131, 125].

| Sample | $\Delta T_{avg}/T_L$ | | $\Delta T_r$ | | $\Delta G^* / k_B T$ | | $K_V$ [m$^{-3}$s$^{-1}$] | |
|---|---|---|---|---|---|---|---|---|
| | EML | ESL | EML | ESL | EML | ESL | EML | ESL |
| Zr, bcc | 0.155 | 0.180 | 330 | 384 | 41.9 | 75.1 | 5.0·10$^{25}$ | 5.0·10$^{41}$ |
| NiZr$_2$ | 0.153 | 0.186 | 213 | 259 | 15.0 | 33.0 | 5.9·10$^{13}$ | 9.7·10$^{21}$ |
| NiZr | | 0.195 | | 300 | | 59.0 | | 1.0·10$^{35}$ |
| CuZr | | 0.244 | | 294 | | 58.0 | | 1.0·10$^{33}$ |

Typical values of nucleation parameters as activation energy $\Delta G^*$ and the pre-factor $K_V$ of nucleation rate are collected in Table 1 for the inter-metallic compounds investigated. It is evident that in all levitation experiments a large undercooling is achieved which is in the order of several hundred degrees, or in terms of relative undercooling ranges from 0.155 to 0.263. For comparison, Turnbull reports values of relative undercooling on the order of 0.18 for a variety of pure metals using the droplet dispersion technique [108]. Concerning pure Zr, a similar value has been obtained by processing a droplet in size of 3mm in the ESL. Even more interesting, a high value of the pre-factor results from our investigations of nucleation statistics, it is on the order of $K_V = 10^{41}$ m$^{-3}$s$^{-1}$. That is more than the value estimated by Turnbull for homogeneous nucleation, but comparable with the value given by Dantzig and Rappaz for homogeneous nucleation. The small half width and the steep fall of the function at the high undercooling side supports the conclusion that the undercooling limit as set by the onset of intrinsic homogeneous nucleation is achieved. We estimate from these experiments the interfacial energy of solid nucleus and undercooled melt. This yields a value of the dimensionless interfacial energy of bcc structured nucleus of $\alpha \approx \sigma/\Delta S_f = 0.61$ with $\sigma$ the interfacial energy and $\Delta S_f$ the entropy of fusion. This value is compatible with estimates [132] from previous undercooling experiments of pure Fe using electromagnetic levitation [133]. Note that Fe nucleates in bcc structure like Zr. It is interesting to compare this value with results of atomic simulations. Even though large undercoolings have been measured approaching a value of relative undercooling $\Delta T_r = \Delta T/T_L \approx 0.18$ as reported by Turnbull for pure metals we estimate a lower limit if the dimensionless interfacial energy, $\alpha \cdot f(\vartheta)^{1/3} = 0.61$ and compare this value with results of atomistic modeling. The corresponding figures are compiled in Table 2

**Table 2:** Results of different calculations of the dimensionless solid-liquid interfacial energy α. DPHS, dense packing of hard spheres; DFT, density functional theory; MD, molecular dynamics simulation; HS, hard sphere potential; AS, adhesive sphere potential; EAM, embedded atom method.

| Authors | Method | Potential | α |
|---|---|---|---|
| Nelson & Spaepen [134] | DPHS | HS | 0.70 |
| Marr & Gast [135] | DFT | HS | 0.48 |
|  |  | AS | 0.46 |
| Hoyt et al. [136] | MD | Fe-EAM | 0.29 |
|  |  | Fe-EAM | 0.32 |
|  |  | Fe-EAM | 0.36 |

All of these values underestimate the interfacial energy assuming the validity of nucleation approach applied in the present work. Only the negentropic model by Saepen is in agreement with neutron and X-ray scattering experiments on levitation undercooled metals. This suggests large undercooling in the case of crystalline phases as observed in the present experiments, or small undercooling in case of quasicrystalline solids, which show similar short-range order both in liquid and solid [137]. There is one interesting investigation on a ternary Ti-Zr-Ni alloy. An icosahedral I-phase was primarily nucleated in direct competition with a crystalline phase. This result supports that the preference of icosahedral short-range order leads to a decrease of the interfacial energy and promotes the nucleation of a quasicrystalline if it is accessible from the view of thermodynamic boundary conditions [9]. From the investigations, which have been described it is directly obvious that short-range order in the undercooled liquid influences crystal nucleation.

## 3.2 Colloidal suspensions

### 3.2.1 Short range order in the liquid state

The interactions in metals and in colloids are both not sensitive to changes of the temperature. Metals and metal alloys are melted by increasing the temperature. By contrast, colloidal crystals cannot be melted this way, since they are confined to the constant volume of the suspending medium and hence cannot expand. Melting therefore is realized mechanically by application of shear [138]. As a side-remark, to realize this shear-melting in atomic systems, meteorite impacts or thermo-nuclear fireworks are required. Obtained either way, the equilibrium fluids of both material classes appear to be remarkably similar. It therefore was an interesting question, whether this resemblance also holds for the deeply undercooled melts at $T \ll T_M$ in the case of metals and $n_P \gg n_{P,M}$ for colloids. The structure factor of metallic systems was measured by elastic neutron or X-ray diffraction using synchrotron radiation. In the colloidal suspensions with their 3–4 orders of magnitudes larger particle sizes, the structure was determined by Ultra Small-Angle X-ray Scattering (USAXS). To analyze the measured structure factor with respect to short-range ordering the same formalism was utilized for metallic and colloidal systems. Colloidal silica particles of 84±6 nm diameter obtained from Stöber synthesis (cf. section 2.2.1) were utilized [59]. From shear modulus measurements [89] their effective particle charge at maximum interaction conditions was found to be independent of $n$ at a value of $Z_{eff}$=340±20. The typical equilibrium phase diagram shows a re-entrant melting as a function of $c_{NaOH}$ with the crystalline region becoming broader with increasing $n$ [15]. An external peristaltic pump shear melts the colloidal suspension. After cessation of shear the suspension is in a meta-stable state of fluid order, equivalent to the undercooled state of a metallic melt.

The USAXS experiments were carried out at the beamline BW4 at HASYLAB (DESY) [102]. The USAXS technique enables measurements of time-resolved diffraction patterns $I(q) = I_o P(q) S(q)$ of colloidal suspensions. The x-ray intensity $I_o$ is measured with a standard Lupolen sample. The particle form factor $P(q)$ is accessible at low $n$ and large $c_{NaOH}$, where particle interactions are sufficiently suppressed and a disordered state [$S(q)$=1] results. $P(q)$ corresponds to the atomic form factor in metallic systems. For the measured intensity $I(q)$, a conventional background and transmission correction was carried out. Figure 18 shows the time evolution of the structure factor of the colloidal system from the non-equilibrium liquid state (lower curve) to the stable solid phase (upper curve) at a particle number density $n$=113 $\mu m^{-3}$. The time interval between each curve corresponds to 7 s due to the integration interval of 3.5 s and the detector readout time of 3.5 s Figure 18 demonstrates that at lower times (< 25 s) there is no change of the structure factor. This assures that especially the first structure factor taken after 3.5 s corresponds to a meta-stable liquid state. For longer times a structural transition into a stable bcc phase is observed. This behavior is representative for the analyzed system over the accessible range of particle concentrations at maximum interaction.

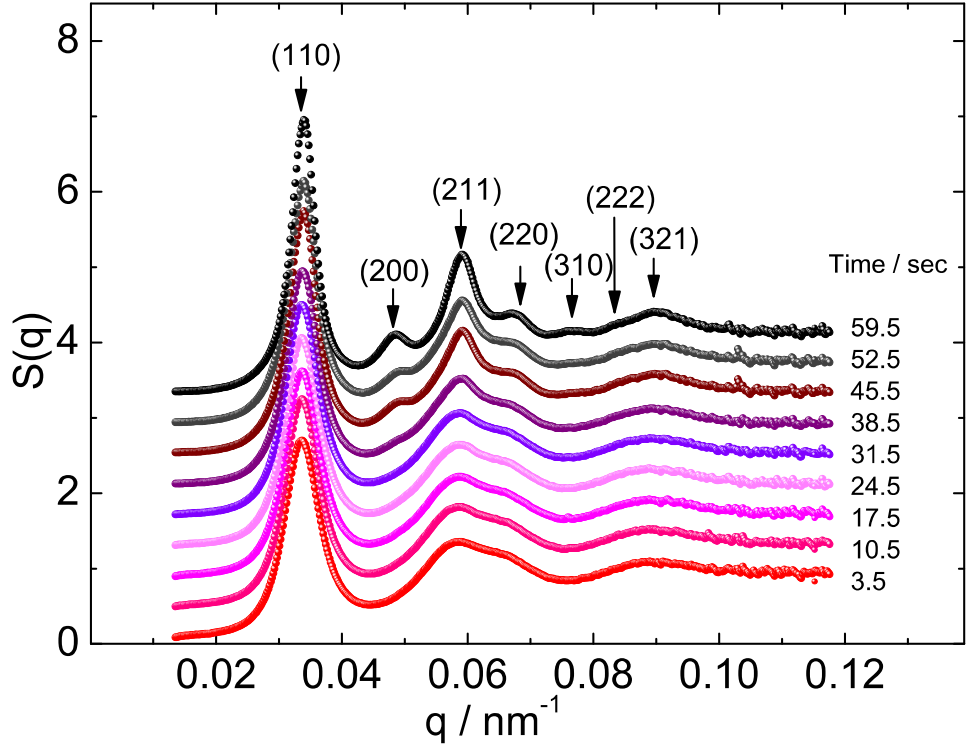

**Figure 18:** Static structure factors $S(q)$ at different times for a colloidal suspension at $n=113$ $\mu m^{-3}$ and maximum interaction that was shear melted (lower curve) and subsequently crystallizes into a bcc structure (upper curve). $S(q)$ are shifted for clarity; time increases from bottom to top. Miller indices are indicated, clearly identifying the bcc structure of the polycrystalline solid formed. (Reprinted with permission from PhD Thesis Work Ina Klassen, *Charged colloids as model systems for metals*, Ruhr-University Bochum (2009)).

Figure 19 compares experimental results for the structure factors of the non-equilibrium fluid state for a colloidal system measured by USAXS (a) with results of the experimentally determined structure factors of undercooled metallic melt on liquid Ni (cf. Figure 9) [15]. In order to compare directly the structure factors of the colloidal suspension and the metal, $S(q/q_{max})$ is plotted as a function of dimensionless wave vectors $q/q_{max}$ with $q_{max}$ the wave vector at the first maximum of $S(q)$. The deviations from equilibrium for the colloidal suspension are given by the chemical potential difference between the meta-stable and stable fluids, $\Delta\mu$. According to Aastuen *et al.* [29]. $\Delta\mu$ is independently determined by measuring the growth velocities of colloidal crystals in dependence on the particle number density $n$ by means of optical microscopy. The growth velocity V obeys a Wilson-Frenkel behavior

$$V = V_0 \left[ 1 - \exp\left(\frac{\Delta\mu}{k_B T}\right) \right] \tag{3.9}$$

where $V_0$ is the limiting velocity and $\Delta\mu/k_B T = B(n-n_L)/n_L$, with the fit parameter $B$ and the particle number density of the fluid in equilibrium with a coexisting crystal $n_L = 13.0$ $\mu m^{-3}$. Corresponding driving forces in metallic systems are determined via the Gibbs free energy difference $\Delta G = G_{solid} - G_{liquid} \propto \Delta\mu$ which is estimated by the linear approximation $\Delta G = \Delta S_f(T_L - T)$ with $\Delta S_f$ the entropy of fusion and $T_L$ the melting temperature.

The diffraction measurements on metal and colloidal suspension are strikingly similar and show up to four oscillations with decreasing intensity for increasing $q/q_{max}$. In both systems, the experiments reveal an asymmetric second oscillation of $S(q/q_{max})$ with a shoulder that becomes more pronounced with increasing deviations from thermodynamic equilibrium. The shoulder position corresponds to the length scale between particles surrounding a central one in a fivefold-symmetric way. The shoulder can therefore be considered as an indication for icosahedral short-range order to be present in both physically different systems [121] and seems to be a generic feature of both material classes.

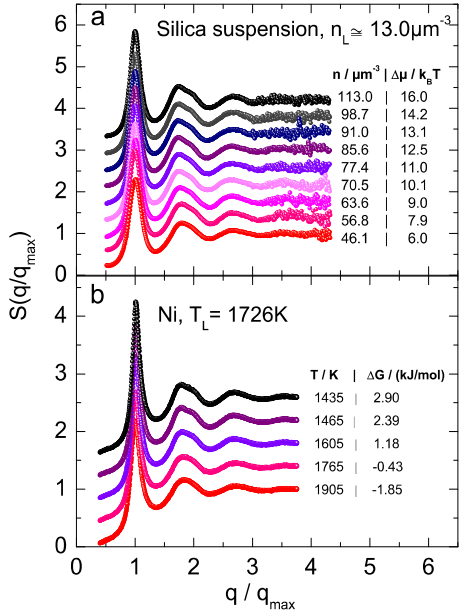
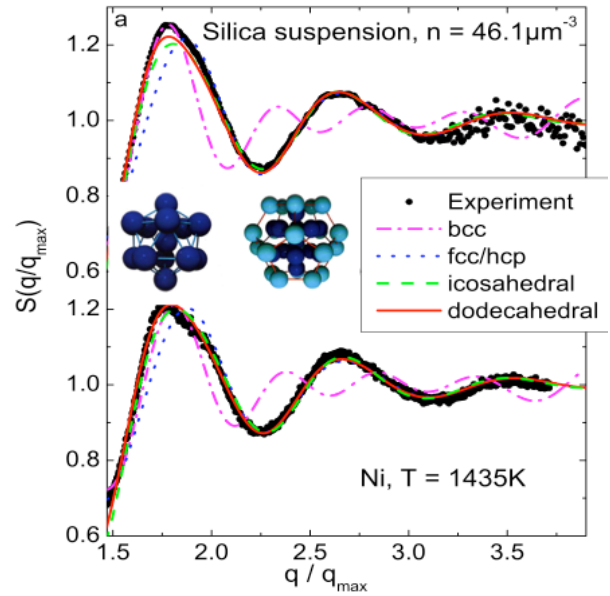

**Figure 19:** Structure factor $S(q/q_{max})$ as a function of normalized scattering vector $q/q_{max}$, measured on the liquid phase of a colloidal suspension at maximum interaction by USAXS (a) and measured on liquid Ni by neutron scattering (b). For the two physically different systems the same behavior is observed. A shoulder appears in the second diffraction peak, which becomes more pronounced with increasing meta-stability, i.e., increasing chemical potential difference for colloids, $\Delta\mu$, and increasing driving force, $\Delta G$, for metals. (Reprinted with permission from PhD Thesis Work Ina Klassen, *Charged colloids as model systems for metals*, Ruhr-University Bochum (2009)).

**Figure 20:** Structure factors $S(q/q_{max})$ of a colloidal melt with $n$=46.1 $\mu m^{-3}$ (a) and a Ni melt with $T$=1435 K {b}. Measured data are represented by symbols and lines show results of simulations assuming short-range order with different symmetries prevailing in the melt: bcc (dash dotted), fcc and hcp (dotted), icosahedral (dashed), and dodecahedral (solid).

The appearance of a shoulder in the second peak of the structure factor S(q) cannot be reproduced by simulations of a simple hard sphere system [139]. Integral equation ap-proaches for hard sphere Yukawa fluids under equilibrium conditions also do not show this feature [140]. According to the approximation proposed by [106] for $S(q/q_{max})$, their simulations assume the liquid to be composed of isolated structural units and neglect intercorrelations of these units. This model is used here for simulation of the structure factor both for metals and colloids. Body centered cubic (bcc), face centered cubic (fcc), or equivalently hexagonal close-packed (hcp) clusters as well as perfect icosahedral and dodecahedral clusters were considered. This certainly neglects the wealth of imperfect clusters of fivefold symmetry [141], which are also likely to be present, but it is sufficient to unequivocally support the conclusions. Using isolated clusters instead of a continuous structure results in low amplitudes of the principal peak. Hence a three-parameter fit to the remaining three oscillations was performed, optimizing the shortest mean distance $r_o$ of the particles, its mean thermal variation $\delta r_o^2$, and the fraction of atoms organized in each structural unit, $c$. The simulated structure factors of the colloidal suspension and the metallic melt are compared with the experimental results in Figure 20. The shoulder in the second peak of the structure factor S(q) resembles that of the molten metal.

Note that a bcc-like short-range order of the undercooled melt utterly fails in describing the data. An fcc-type short-range order performs better but does not agree with the shape and peak positions in the second oscillation, where the significant asymmetry of a shoulder is observed. An agreement is achieved, if icosahedral units of fivefold symmetry are assumed. It becomes even better, if larger dodecahedral aggregates are assumed. For the colloidal suspension, the best fit of the simulations to the experiments for icosahedral ordering gives the values: $r_o = 315$ nm, $\delta r_o^2 = 487$ nm$^2$, and $c = 99\%$. A dodecahedral type of ordering shows the related parameters as $r_o=319$ nm, $\delta r_o^2 = 487$ nm$^2$, and $c = 99\%$. By comparison, for the simulation of liquid Ni $r_o = 0.238$ nm, $\delta r_o^2 = 0.29 \cdot 10^{-3}$ nm$^2$, and $c = 99\%$ (icosahedral), and $r_o=0.242$ nm, $\delta r_o^2 = 0.26 \cdot 10^{-3}$ nm$^2$, and $c=95\%$ (dodecahedral) are obtained.

The comparison of the structure factors $S(q/q_{max})$ of a colloidal melt and of molten Ni sample shows strikingly similar results if scaled with the scattering vector corresponding to the nearest neighbor distance. In particular, for both these physically different systems an asymmetric shoulder in the second maximum of $S(q/q_{max})$ is observed, which becomes more pronounced with increasing deviations from equilibrium. Hence, low-salt hard core Yukawa interactions in concentrated colloidal systems are found suitable to model LJ-like interactions in metallic systems. Moreover, the experimental data were analyzed by an approach assuming isolated structural units in the liquid. The simulations demonstrate that the assumption of bcc, fcc, or hcp structural units alone does not fully describe the experimental results. The agreement between simulation and experiments becomes nearly quantitative when icosahedral or dodecahedral clusters are assumed. Given the presence of this striking similarity despite the different types of interactions in the two physically different systems, the slightly soft repulsive part of both the LJ-like potential of liquid metals and the hard core Yukawa potential of the charged spheres is a precondition for the formation of icosahedral short-range. The finding that the repulsive term of the interaction potential is decisive for the formation of the short-range order in liquids is in full agreement with previous theoretical investigations [142]. Since electronic effects are absent in colloidal systems, the similarity of the structure factors of the colloidal and the metallic melts suggests that short-range order is controlled by topological effects in both these physically different systems. This finding is of general importance since short-range order essentially governs the formation of solid-liquid interfaces and crystal nucleation in undercooled melts [10].

### 3.2.2 Determination of the metastability of colloidal suspensions in liquid state

The metastabilty of liquid metals is characterized by the amount of undercooling below the melting temperature. Experiments of crystallization in colloidal suspensions are conducted usually under isothermal conditions. Therefore, the difference of the chemical potential between solid and liquid phase is taken to measure the meta-stability of colloidal systems in liquid state. For hard spheres and attractive hard spheres, accurate analytical calculations and simulations providing this quantity are available [143]. For charged spheres, this is not the general case due to the much larger parameter space [144]. $\Delta\mu$ however can be conveniently determined experimentally [145]. At low meta-stability heterogeneous nucleation at the walls of the sample cell dominates and the growth velocity of colloidal crystal can be determined using time-resolved microscopy. After cessation of shear flow, a planar front of twinned bcc crystals propagates linearly with their densest packed planes parallel to the cell wall [146]. Crystallization is monitored by Bragg- or polarization microscopy [145, 147, 148, 149, 150] in a flat flow through cell made of quartz glass with a wall-to-wall distance of 1mm. The cell is mounted on the stage of an inverted microscope equipped with a low-resolution objective. Images are recorded by a charged-coupled device (CCD) camera and stored in a PC for image analysis. Figure 20 (left) shows microscopic images of the crystallization of a colloidal silica suspension (Si84) at low particle number density n=19.0 $\mu m^{-3}$ taken by polarization microscopy in side view of the sample cell at different time intervals. The colored regions represent the crystals growing in planar morphology from both sides of the cell towards the interior. The image size is of 1.00 x 1.16 $mm^2$. The corresponding growth velocity is determined to be 9.7$\mu m/s$. The growth velocity was measured as a function of particle concentration and of the amount of added NaOH. In all cases a linear increase of the crystal dimension with time was observed indicating reaction controlled growth.

At high particle interaction between the particles, homogeneous nucleation dominates and an isotropic polycrystalline material is produced. A typical image of a polycrystalline sample crystallized from a shear melted silica suspension at n=32.0$\mu m^{-3}$ is shown in Figure 20 (right). The crystals appear as faceted and irregularly shaped polyeders of different colors. The facets result from crystal intersections occurring during growth [151]. The color differences originate from different crystal orientations.

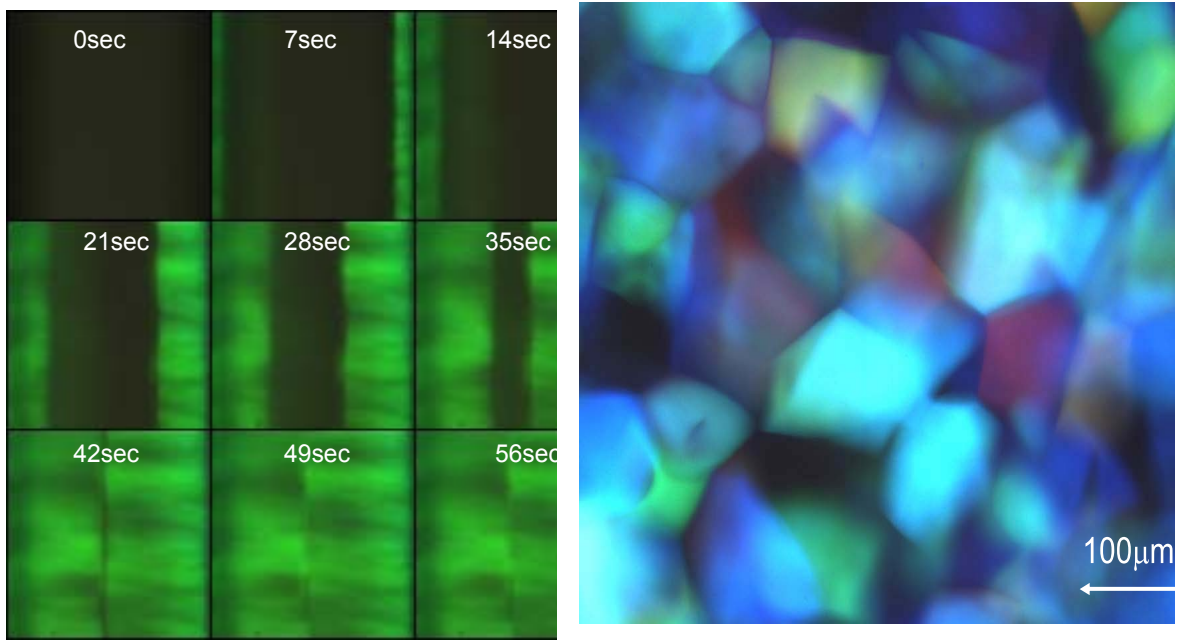

**Figure 21:** (left) Growth of heterogeneously nucleated wall crystals of a colloidal silica suspension with particles of 84nm at particle number density n=19.0μm$^{-3}$ at various time intervals starting from the top to the bottom, (right), polycrystalline solid of a silica system homogeneously nucleated within the bulk of a shear melted suspension at a particle number density n=26.0μm$^{-3}$. The different colors are caused by light scattering from different orientation of the crystallites. (Reprinted with permission from PhD Thesis Work Ina Klassen, *Charged colloids as model systems for metals*, Ruhr-University Bochum (2009)).

Figure 22 shows the growth velocity at maximum particle charge as a function of the particle number density *n*. Firstly, the growth velocity steeply rises with *n* before it approaches a saturation level at large particle number density. Such behavior is well described by the Wilson-Frenkel law following the formalism of Aastuen et al. [29]. Based upon investigations of Palberg [30], Stipp [152] and Brougthon [153] the resulting growth velocity is described in analogy to eq. (3.9) by the equation

$$v = f_0 d_i \cdot \frac{D}{d_{NN}^2} \left[1 - \exp\left(-\frac{\Delta\mu}{k_B T}\right)\right] \qquad (3.10)$$

with $d_i$ the thickness of the interfacial layer, $f_o$ the success rate of particle attachment from the liquid impinge at the solid, $D$ the long time self-diffusion coefficient, $d_{NN}$ the nearest neighbour distance as the characteristic length scale for particle diffusion, $\Delta\mu=\mu_L-\mu_S$ the difference of chemical potential in liquid, $\mu_L$, and in solid, $\mu_S$, state in equilibrium. $k_B$ is Boltzmann's constant and *T* the temperature. Interfacial thicknesses in a one component system are on the order of 3-6 $d_{NN}$ but the thickness drops if small quantities of a second component are added as impurities, and in binary mixtures of 1:1 composition, a monolyer interface was found [154]. The impingement factor $f_o$ is $f_o=1$ for pure metals with more or less isotropic metallic bonding [155], but much less $f_o \approx 0.01$ for pure semiconductors with strongly directional covalent bonding [156]. Since close to the melting point, the fluid structure is very close to that of the crystal [22, 157], it is assumed to be unity for colloids. The pre-factor of eq. (3.9) has the dimension of a velocity and corresponds to the velocity $v_\infty$ of the advancing solid-liquid interface at infinite driving force $\Delta\mu=\infty$. To adapt this formalism to colloidal systems with variable charge and electrolyte content, an assumption for the chemical potential difference is necessary. According to Würth et al. [145] $\Delta\mu$ can be expressed in terms of a rescaled energy density

$$\Delta\mu = k_B T \cdot B \cdot \Pi^* \quad \text{with} \quad \Pi^* = (\Pi - \Pi_F)/\Pi_F$$

if one further assumes $\Delta\mu_C^M \propto \Delta\mu_F^M$ (where M denotes melt, C denotes crystal and F denotes the fluid at freezing).

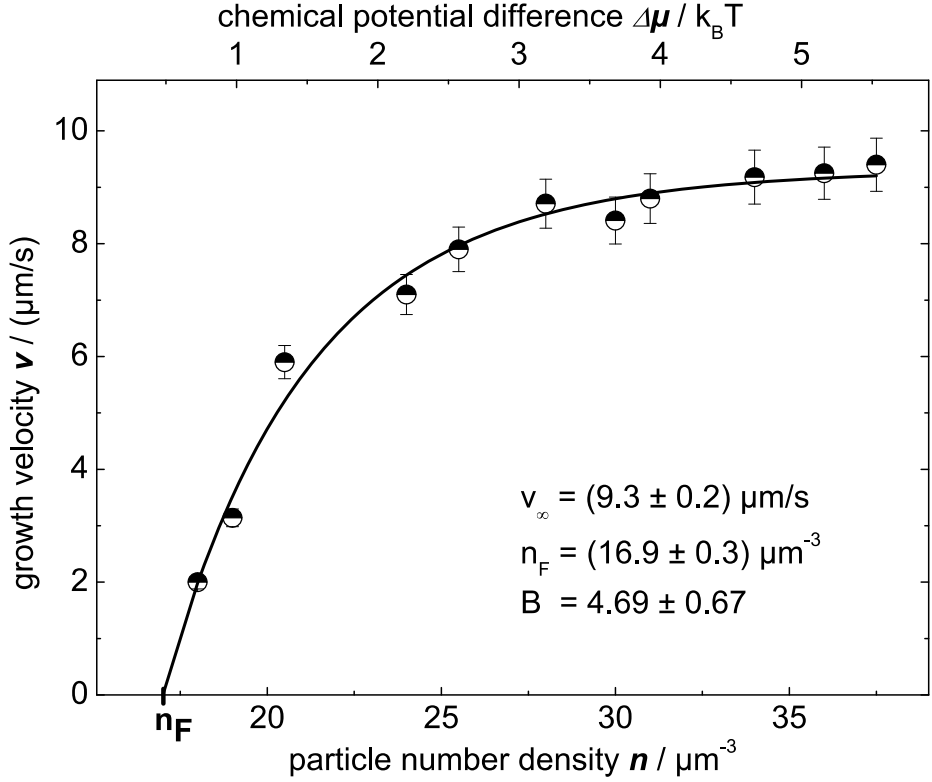

**Figure 22:** The growth velocity v for the Si84 system as a function of the particle number density n at maximum interaction. The solid curve is a fit of a Wilson-Frenkel law according to eq. (3.9) using Würth's approximation for $\Delta\mu$. The fit returns $v_\infty = (9.3\pm0.2)\mu m/s$ and $B=(4.69\pm0.67) k_BT$. The chemical potential difference $\Delta\mu$ between solid and liquid used in the upper x axes is derived using the conversion factor $B$. (Reprinted with permission from PhD Thesis Work Ina Klassen, *Charged colloids as model systems for metals*, Ruhr-University Bochum (2009)).

The energy density is given as $\Pi = AnV(d_{NN})$, with $A$ being the particle coordination number in the melt and the equilibrium fluid, $V(d_{NN})$ is the interaction energy at nearest neighbor distance. This approach considers both the direct linear density dependence of $\Delta\mu$ and the pair interaction energy $V(r)$. $B$ is a proportionality constant that is determined from experiment as fitting parameter. The best fit to the data of Figure 22 yields $v_\infty = (9.3\pm0.2)\mu m/s$ and $B=(4.69\pm0.67) k_BT$ with $T = 298$ K. Equation (3.10) describes the experimental data well with the fitting parameters as given above. The difference of the chemical potential between liquid and solid state as a function of the particle number density shows a nearly linear dependence of $\Delta\mu$ on $n$, which can be extrapolated at higher particle number densities, where the growth velocity is more difficult to determine due to the preference of homogeneous nucleation [158].

### 3.2.2 Nucleation studies in colloidal suspensions

In metallic systems the formation of nuclei in undercooled melts is very fast. The frequency of impingement of atoms from a liquid to a growing nucleus is very high and in the order of $10^{13}$ Hz. This makes it extremely difficult to observe directly nucleation processes in undercooled metallic melts. Colloidal systems are frequently discussed as model systems to study nucleation phenomena. In contrast to atomic systems they provide a sufficiently long experimental time scale for investigations. This is due to the large size of their particles (hundreds of nm) and the Brownian nature of their motion. Therefore, time dependent investigations of nucleation processes are possible. Nucleation was in situ observed in three dimensional real space in colloidal hard and soft sphere suspensions by applying confocal microscopy [42, 43, 159, 160]. This allowed for studies with 'atomic' resolution and revealed that the actual nuclei are spherical on average. Individual nuclei are both irregular and show a fuzzy surface at early stages. This has been confirmed by simulations of both hard [161] and charged [162] spheres. Anisometry and fuzzyness are also observed in metals e.g. in a recent combined study of experiments on Ni and the corresponding simulations [163]. Later stages of nucleation can also be observed by video microscopy with lower resolution. They typically show near spherical crystallites [29, 150]. Only rarely, dendritic growth has been reported [164, 165]. Still classical nucleation theory can be used for interpreting scattering experiments, as these involve an orientational averaging over a large number of nuclei or crystals.

An important point is that heterogeneous nucleation prevails only at very low meta-stability [95]. To be studied at all, it has to be deliberately introduced by seeds [38]. Without such measures, homogeneous nucleation prevails and the evolution of the crystallites from the meta-stable fluid state into the stable solid state can be followed by different time resolved methods. The nucleation rate density $J(n)$ varies drastically with increasing n. Therefore, different techniques must be employed in studies covering a large range of meta-stabilities. At low densities, $J(n)$ is determined from video-microscopy, at medium densities via post-solidification size counts from polarization microscopy, and at high densities from Bragg scattering or time resolved USAXS measurements [15, 150, 166]. All three approaches yield consistent $J(n)$ for a given sample [158].

In scattering experiments, the obtained $S(q,t)$ contains contributions of both crystals and melt. The structure factor of the crystalline phase can be extracted following the method of Harland and van Megen [167]. The melt background $S_M(q)$ has to be scaled so that the intensity at the minimum q' of the structure factor is equal to the intensity recorded for the following time scans $S_M(q',t) = \beta(t) \cdot S_f(q')$ with a scale factor $\beta(t) \cdot$ chosen at each time step. This scaling assumes that the fluid density and composition retain the same throughout the crystallization. The crystalline part of the structure factor is determined by subtracting the scaled fluid part from the measured structure factor

$$S_c(q,t) = S(q,t) - \beta(t) \cdot S_f(q) \tag{3.11}$$

where $\beta(t)$ is the time dependent melt fraction. This scaling assumes that the underlying structure factor of liquid will not change its shape during crystallization: the density and composition of the fluid retain the same throughout the crystallization. The resulting crys-tal structure factor $S_c(q,t)$ is used to determine the properties characterizing the crystallization kinetics. The integrated intensity is related to the crystal fraction called crystallinity, $X(t)$, the peak position $q_m$ to the lattice spacing of the crystals and the full width at half height of the peak width to the average size of the crystallites, L(t) [168].

The crystallinity, $X(t)$, is determined by integrating the structure factor over the area of the main Bragg reflection:

$$X(t) = c \int S_{xtal}(q,t) \, dq \tag{3.12}$$

The lattice constant $g$ of the crystal phase is calculated from the peak position $q_{hkl}$

$$g(t) = \frac{2\pi}{q_{hkl}(t)} \sqrt{h^2 + k^2 + l^2} \tag{3.13}$$

where h, k, l are the Miller indices.

The average crystal size $L$ is inverse to the full width at half height $\Delta q$ following the Scherrer equation:

$$\langle L(t) \rangle = \frac{2\pi K}{\Delta q(t)} \tag{3.14}$$

From these basic parameters, the number density of crystallites $n_{xtal}$ and the nucleation rate density $J(t)$ can be determined. The number density of crystallites is deduced from

$$n_{xtal}(t) = \frac{X(t)}{\langle L^3(t) \rangle} = \frac{X(t)}{\alpha \langle L(t) \rangle^3} \tag{3.15}$$

with the parameter $\alpha \approx 1.25$ relating the average crystal size cubed [169]. The nucleation rate density is defined as the rate at which crystals appear in the remaining melt volume:

$$J(t) = \frac{1}{(1-X(t))} \cdot \frac{d}{dt} \frac{X(t)}{\langle L^3(t) \rangle} \tag{3.16}$$

This quantity represents the number of critical nuclei, which form inside a unit volume. The nucleation rate is normalized with respect to the remaining liquid volume, 1-X(t).

Figure 23 illustrates the time dependent behavior for the parameters extracted from the sequence of the time dependent structure factor measured: (a) crystallinity $X$, (b) average crystal size $L$, (c) crystallite density and (d) nucleation rate densities $J$. The typical crystallization experiment in charged colloids shows a sigmoidal curve of crystallinity versus time as shown in Figure 22. The curve exhibits initially a sharp increase of the crystallinity, which is attributed to nucleation and growth of crystals. Once the sample has reached the equilibrium state, this process decreases. The slow rise at long times is identified as ripening or coarsening process where large crystals grow at the expense of smaller ones. The crystal of average size L grows to a size of approximately 1.1 µm and the nucleation rate density first increases from $J = 10^{16} m^{-3} s^{-1}$ by about one to two orders of magnitude, achieves a maximum value at $J = 8 \cdot 10^{17} m^{-3} s^{-1}$ and decreases again. The parameters introduced are important to understand and to describe nucleation processes in colloids as well as in atomic or molecular systems. The time dependence of the nucleation parameters is not accessible in metallic systems with the exception of crystallization experiments of me-tallic glasses [170, 171]. In crystallization experiments of metallic glasses the time evolution of nucleation is controlled by the high viscosity of the deeply undercooled melt slightly above the glass transition temperature. It is very sluggish opposite to nucleation processes in melts undercooled below their melting temperature in a temperature range, where the viscosity is much smaller than in the region around the glass transition temperature.

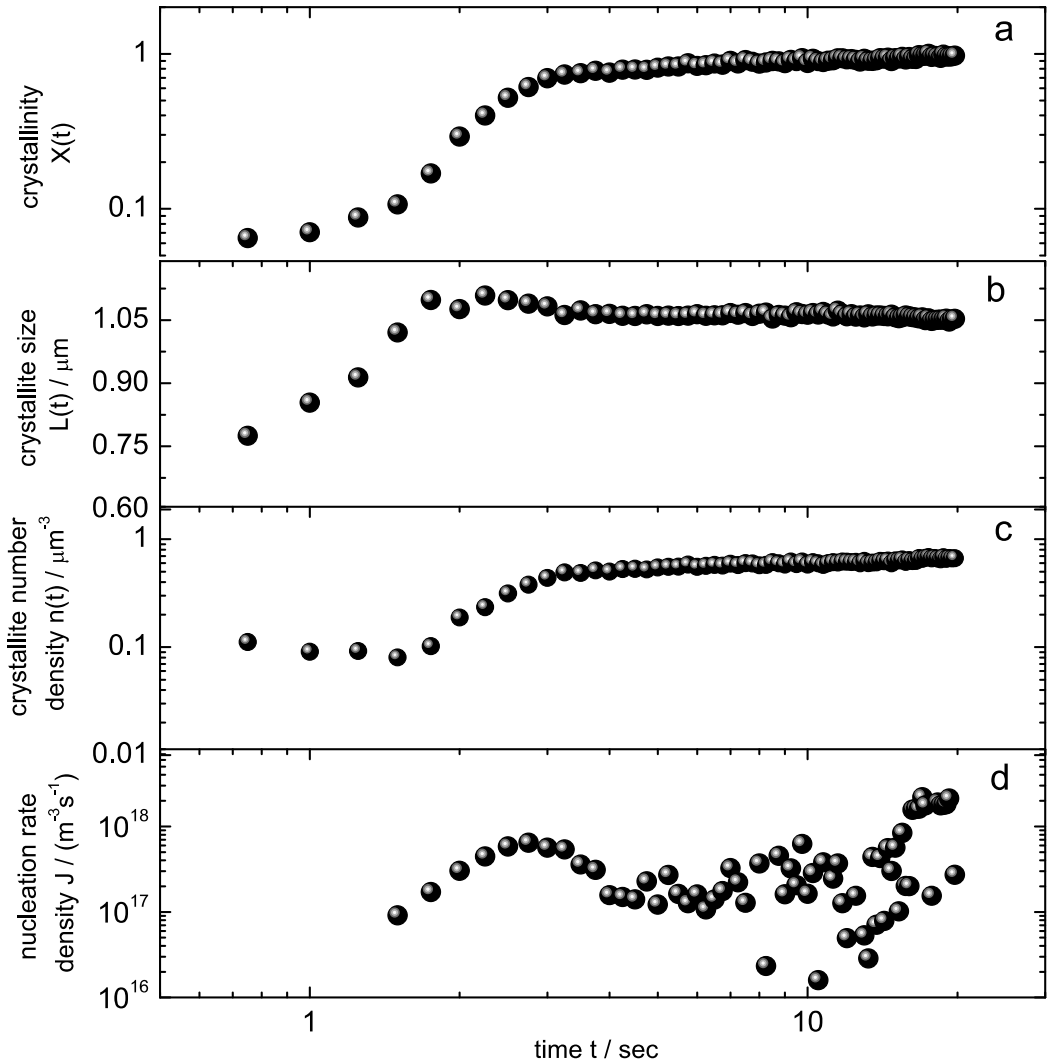

**Figure 23:** Time traces of extracted nucleation parameters from scattering data (a) crystalline volume fraction X(t) or crystallinity, (b) average crystallite size L(t), (c) crystallite number density n(t), and (d) nucleation rate densities *J(t)* exemplarily shown for a charged silica system with particles of size 77nm and a particle number density n=224.7μm$^{-3}$. (Reprinted with permission from PhD Thesis Work Ina Klassen, *Charged colloids as model systems for metals*, Ruhr-University Bochum (2009)).

Crystal nucleation is considered a thermally activated process with competing bulk and surface contributions to its energy balance. The nucleation rate is obtained by considering the transition rates of elementary particles from the fluid to the growing nucleus and vice versa. By considering the Boltzmann distribution function for classical systems (no quantum physical processes) the nucleation rate density increases exponentially with increasing deviation from thermodynamic equilibrium. According to classical nucleation theory (CNT) [172] clusters of crystalline structure are formed in statistically independent events by stepwise addition of particles from the fluid, The formation of clusters of critical size is thermally activated across a barrier due to the interfacial energy between crystal nucleus and surrounding liquid, $\gamma > 0$. The nucleation barrier $\Delta G^*$ to form sphere like nuclei of critical size of colloidal suspensions in fluid state is given by

$$\Delta G^* = \frac{16\pi}{3} \cdot \frac{\gamma^3}{(n\Delta\mu)^2} \tag{3.17}$$

where $n$ is the particle number density and $\Delta\mu$ the difference of chemical potential between solid and liquid state. For colloidal particles crossing the solid-liquid interface, the steady state nucleation rate, $J_{ss}$, is determined by

$$J_{ss} = J_o \exp\left(-\frac{\Delta G^*}{k_B T}\right) \quad \text{with} \quad J_o = 12\left(\frac{4}{3}\right)^{2/3} \pi^{-1/3} n^{4/3} \sqrt{\frac{\gamma}{k_B T}} D_L^S \tag{3.18}$$

where $J_o$ is the kinetic pre-factor that was explicitly calculated for colloidal suspensions [158] within the framework of classical nucleation theory [173]. $D_L^S$ denotes the long-time self-diffusion coefficient. Despite its simplicity, the CNT is widely used to parameterize nucleation both in metallic undercooled melts [173] and colloidal systems as well [30, 34, 36, 37]. While in metallic systems the interaction potential acting on the individual atoms is fixed, the interaction potential in charged colloidal suspensions depends on the particle number density $n$ or volume fraction $\Phi$.

In systems with high relaxation rates the formation of clusters within the liquid follows instantaneously any change of state of the systems. Under such circumstances nucleation is considered as taking place under steady state conditions. Such conditions are certainly present in liquid metals in a temperature range around the melting temperature in which atomic relaxation takes place very rapidly because of very high self-diffusion coefficients in the liquid state. In monoatomic metallic systems atomic movement in the liquid is even considered to be collision limited i. e. the frequency in the order of the Debye frequency will set the limit for the frequency of atomic place changes [174]. Even in melts of alloys, in which atomic movement is diffusion controlled atomic replacement in liquid state is occurring rapidly compared to the time scale of changes in their state for instance during rapid cooling [175]. On the other hand during crystallization of metallic glasses around the glass transition temperature at which the viscosity is very high in the order of $10^{13}$ Poise compared to about $10^{-2}$ Poise at the melting temperature of pure metals, transient effects become important in crystal nucleation processes in metallic glasses [170].

In colloidal suspensions movement of the particles in liquid state is controlled by Brownian motion. Compared with movements of atoms in liquid metals, Brownian motion is very sluggish. Therefore, one would expect that transient effects in crystallization of colloids could be of importance. To consider transient effects, the experimentally determined time-dependent nucleation rate densities were evaluated applying the theory of transient nucleation by Kashchiev [176]. Accordingly, the time dependent nucleation rate is given by:

$$J(t) = J_{ss}\left[1 + 2\sum_{m=1}^{\infty}(-1)^m \cdot \exp\left(-\frac{m^2 t}{t_i}\right)\right] \tag{3.19}$$

where $J_{ss}$ denotes the steady state nucleation rate and $t_i$ the induction time given by

$$t_i = \frac{\gamma d_{NN}^2 k_B T}{\left(\frac{3}{8}\right)\left(\frac{4}{3}\right)^{2/3} \pi^{5/3} D_L^s n^{2/3} (\Delta \mu)^2} \tag{3.20}$$

The induction time determines the time of the delay for the onset of crystal nucleation in experiments of rapid changes of the states.

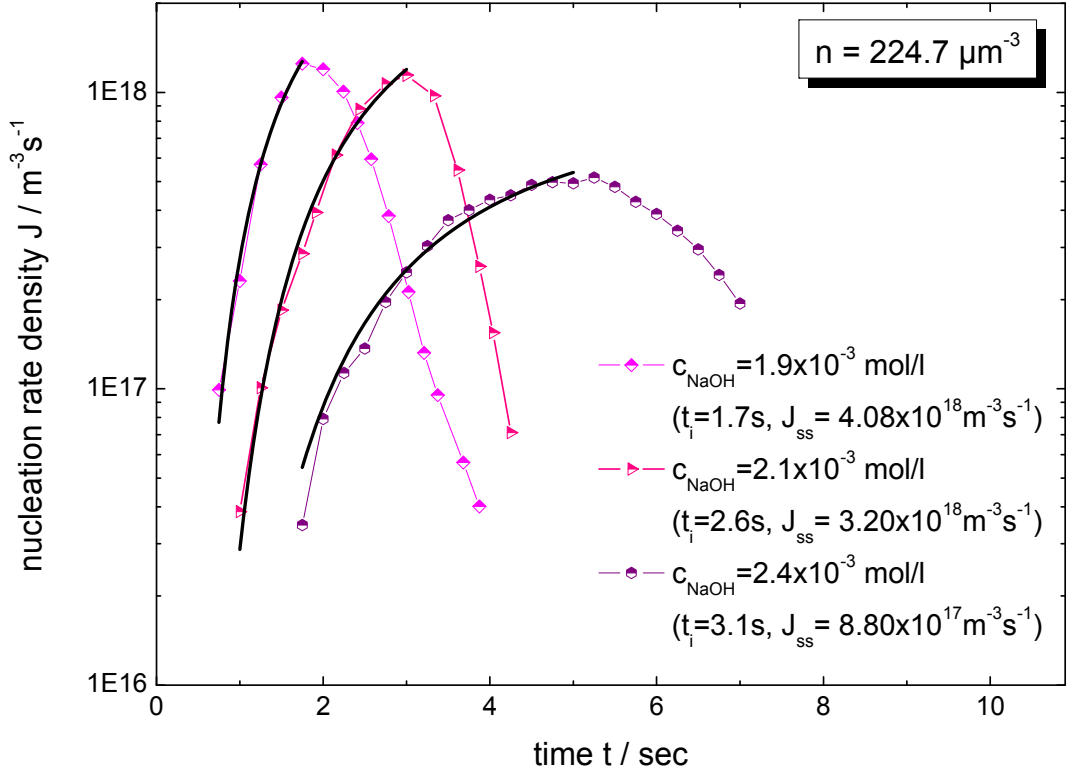

**Figure 24**: Steady state nucleation rate densities $J_{ss}$ and the corresponding induction time $t_i$ obtained by fitting the measured time dependent nucleation rate density $J(t)$ of silica colloidal suspensions at fixed particle number density n = 224.7 μm$^{-3}$ according to the expression developed by Kashchiev for transient nucleation. (Reprinted with permission from PhD Thesis Work Ina Klassen, *Charged colloids as model systems for metals*, Ruhr-University Bochum (2009)).

There are two unknown parameters in eq. (3.20). That is the long time self-diffusion coefficient $D_L^s$ and the interfacial energy $\gamma$. The regard of transient effects in nucleation of colloidal suspensions offers the possibility to determine independently the self-diffusion coefficient $D_L^s$ from the measured transient time according to eq. (3.20). In colloidal suspensions, the self-diffusion coefficient will depend on the particle number density, $D(n)$. Thus, the interfacial energy, $\gamma$, can be determined from measurements of nucleation rates provided homogeneous nucleation dominates. Figure 23 shows measured nucleation rates for silica colloidal suspensions at fixed particle number density n=224.7μm$^{-3}$ at three different salty concentrations.

No solution of eqs. (3.18) and (3.20) does exists for the nucleation rates $J(t)$ shown in Figure 24. According to Figure 23 for all investigated silica colloidal suspension the nucleation rate steeply rises passes through a maximum and rapidly falls at large times. The nucleation rate density does not approach a stationary state where a dynamic equilibrium is achieved in the formation of nuclei per volume and time unit. We assume that the nucleation rate collapses without achieving the steady state nucleation rate. This assumption is supported by a comparison of the results of silica suspension with investigations of nucleation behavior in polystyrene colloidal suspensions [150]. Figure 25 displays the time dependent nucleation rates of a completely deionized polystyrene colloidal suspension (PnBAPS68) with particles in size of 68nm±3nm, which are dispersed in water. These particles carry strongly acidic sulfate groups on their surface, which are completely dissociated in an aqueous environment leading to a maximum effective charge. To increase the particle interaction, the particle number density was slightly increased from 18 $\mu m^{-3}$ up to 20 $\mu m^{-3}$. One can nicely see the transition from a nucleation behavior reaching a steady state of nucleation at n = 18 $\mu m^{-3}$ to a transient nucleation behavior at n = 20 $\mu m^{-3}$. The lines again are fits of Kashchiev's theory. At n = 18 $\mu m^{-3}$ the plateau in J(t) and $J_{SS}$ agree nearly quantitatively, while $J_{SS}$ is much larger than the maximum J(t) for n = 20 $\mu m^{-3}$. This situation clearly parallels the findings on J(t) in the highly concentrated Si77 system. However, in both cases the steady state nucleation rates inferred from experiments are far below (between 6 and ten orders of magnitude) than $J_o$ predicted by CNT.

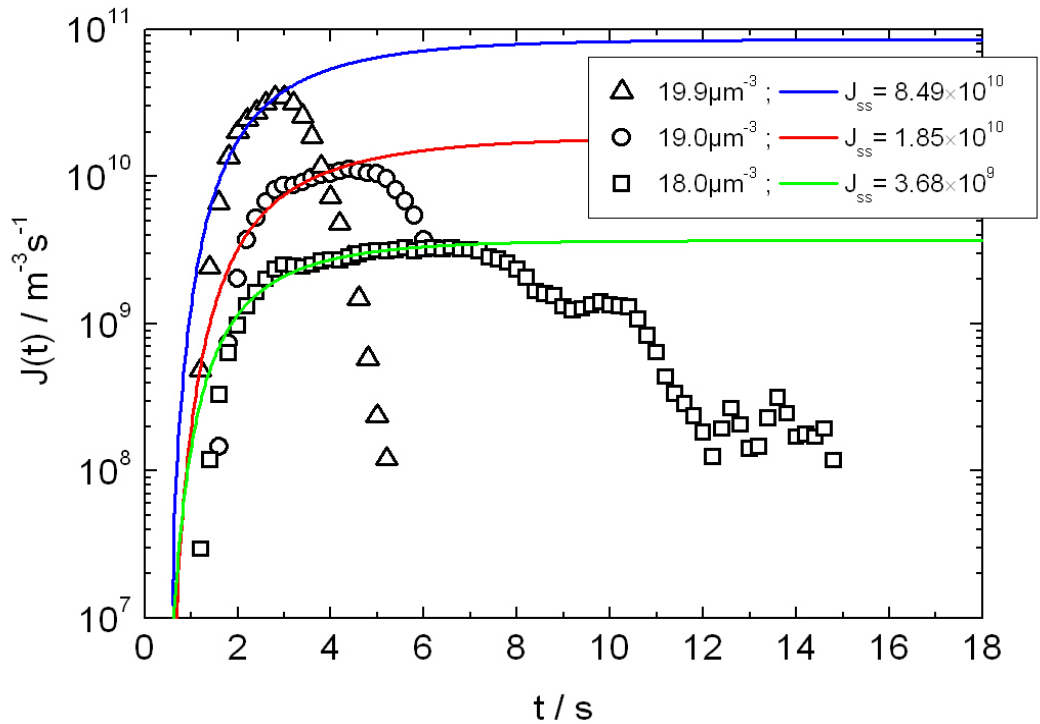

**Figure 25:** Time dependent nucleation rate *J(t)* as measured on polystyrol colloidal suspensions PnBAPS68 at various particle number densities n=19.9$\mu m^{-3}$ (blue), n=19.0$\mu m^{-3}$ (red), and n=18.0$\mu m^{-3}$ (green) [150]. (Reproduced from P. Wette, I. Klassen, D. Holland-Moritz, T. Palberg, S. V. Roth, D. M. Herlach*, Colloids as model systems for liquid undercooled metals,* Physical Review E, Rapid Communication **79** (2009) 010501(R) with the permission of APS Publishing, Copyright 2009 APS).

There are two unknown parameters in eq. (3.20). That is the long time self-diffusion coefficient $D_L^S$ and the interfacial energy $\gamma$. The regard of transient effects in nucleation of colloidal suspensions offers the possibility to determine independently the self-diffusion coefficient $D_L^S$ from the measured transient time according to eq. (3.20). In colloidal suspensions, the self-diffusion coefficient will depend on the particle number density, $D(n)$. Thus, the interfacial energy, $\gamma$, can be determined from measurements of nucleation rates provided homogeneous nucleation dominates. Figure 23 shows measured nucleation rates for silica colloidal suspensions at fixed particle number density n=224.7μm$^{-3}$ at three different salty concentrations.

The phase behavior of a bi-disperse sample consisting of charged colloidal silica spheres with diameters $d_A$ = (104.7 ±9.0) nm, $d_B$ = (88.1 ± 7.8) nm, and a composition $A_{0.6}B_{0.4}$ as well as the time-resolved homogeneous nucleation of colloidal polycrystals from their icosahedral short-range order were investigated by USAXS experiments. At particle number densities below n ≈ 100 μm$^{-3}$ a similar phase behavior is observed as for monodisperse, single component sample. However, at particle number densities above n ≈ 100 μm$^{-3}$ and below the maximum achieved n = 201 μm$^{-3}$ a change in the phase behavior is observed, resulting in – yet unexplained – (i) shifts of the phase boundaries to higher sodium hydroxide concentration and (ii) a bifurcation in the qualitative crystallization behavior, with respect to the specific emergence of the obtained bcc-type crystal structure. Quantitatively, a solid-liquid interfacial energy in the order of μJ/m$^2$ and a Turnbull coefficient of $C_T$ = 0.267 are estimated [177].

### 3.2.3 Interfacial energy, entropy and enthalpy of fusion

The interfacial energy is the barrier for the formation of a nucleus of critical size and therefore one of the most important parameters in nucleation and phase transformations of first order. From the combined data sets of $J(n)$ and $\Delta\mu(n)$ the nucleus-melt interfacial energies $\gamma(n)$ are derived within the framework of classical nucleation theory. Either a least-square fit of eq. (3.16) to the experimental data using a proportionality constant A, an effective long time self-diffusion coefficient $D_L^S(n)$ and $\gamma(n)$ as free parameters, or a graphical evaluation method can be utilized in which $\gamma(n)$ is determined from the local slope without any assumption about the kinetic pre-factor. More details are given in [166, 158].

The absolute values of the interfacial energy are smaller by orders of magnitude for colloids than for metals. This is due to the different particle number densities, n, involved for e.g. metals (n ≈ 10$^{26}$m$^{-3}$) and colloids ((n ≈ (10$^{17}$- 10$^{19}$)m$^{-3}$)) $(n \approx (10^{17} - 10^{19}) m^{-3})$. In view of this we follow the work by Turnbull [108] and normalize $\gamma$ with the area taken by a single particle in the interface, $A_p$, to compare reduced values of the interfacial energy, $\sigma = \gamma A_p$ $\sigma = \gamma A_p$ of the various physical systems. Different measures for $A_p$ have been proposed. For metals [108] and hard spheres [178, 179, 180], but also for strongly screened charged spheres [181] $A_p$ was approximated by $A_P \approx (2a)^2$ where $a$ is the particle radius. This is reasonable since at the large volume fractions encountered in close-packed metals, hard spheres, and slightly charged hard sphere crystals, the particles are nearly in contact. By contrast, in low salt charged sphere crystals the nearest neighbor distance at melting is usually on the order of several particle diameters due to mutual electrostatic repulsion.

Here, the area of interest is the square of the nearest neighbor distance, $A_P \approx d_{NN}^2$. The spread in $d_{NN}^2$ covers about three orders of magnitude between its value at melting for PnBAPS70 ($n_{M,PnBAPS70}=2\mu m^{-3}$) and the one at the largest particle concentration for Si77 ($n_{M,Si77}=80\mu m^{-3}$). Therefore, each interfacial energy is normalized by the square of the nearest neighbor distances at the particle number density investigated, $A_P \approx d_{NN}^2$; $d_{NN}^2 = n^{-2/3}$. As the energy unit for the reduced interfacial energy, $k_B T_m$ is used with $T_m$ the melting temperature.

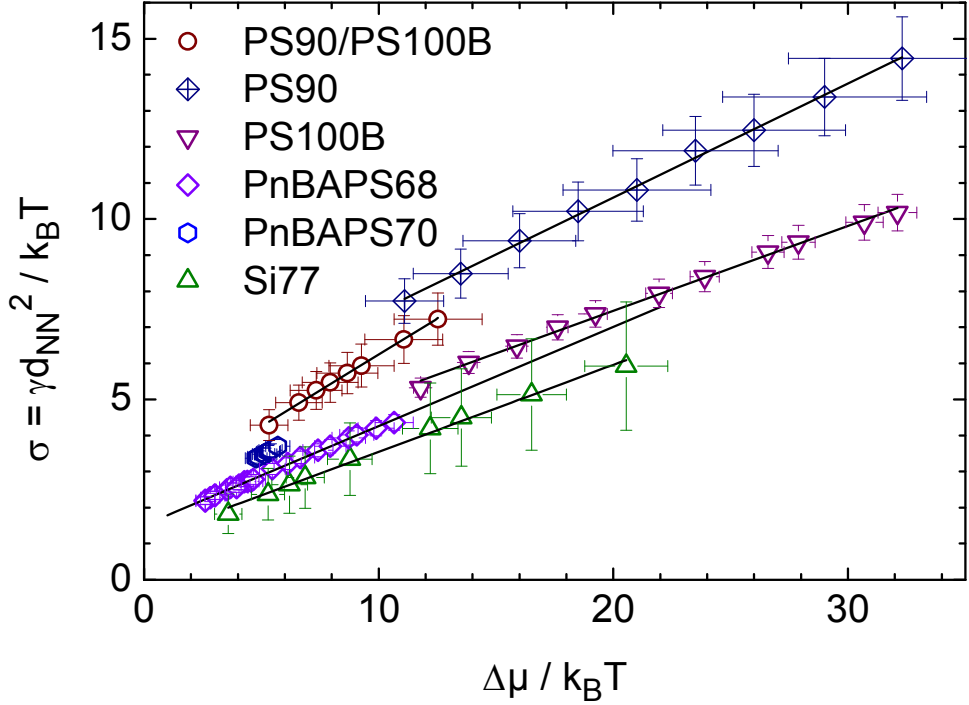

**Figure 26:** Dependence of the reduced interfacial energies on the metastability $\Delta\mu$ for the indicated species. Interfacial free energies $\gamma$ (in J/m$^2$) were normalized by $n^{-2/3}$ and plotted in units of $k_B T$ as quoted from the original literature. Solid lines are least-square fits of $\sigma = \sigma_o + m\,\Delta\mu$. (Reproduced from T. Palberg, P. Wette, D.M. Herlach, *Equilibrium fluid-crystal interfacial free energy of bcc-crystallizing aqueus suspensions of polydisperse charged spheres*, Physical Review E **93** (2016) 022601, with the permission of APS Publishing, Copyright 2016 APS).

In Figure 26 the data of the reduced interfacial energy $\sigma = \gamma n^{-2/3}/k_B T$ is plotted as a function of the metastability $\Delta\mu$ for PnBAPS68, PnBAPS70, Si77, PS100B, PS90 and the mixture of PS90 and PS100B. It is obvious that the reduced interfacial energy of all charged sphere samples show a linear increase for increasing metastability. A scheme according to Turnbull's rule [173]

$$\sigma_o = C_T \frac{\Delta H_f}{N_A} = C_T T_m \frac{\Delta S_f}{N_A}$$

is used to obtain the values of the interfacial energy $\sigma_o$ extrapolated to $\Delta\mu = 0$, the entropy of fusion $\Delta S_f$ and the heat of fusion $\Delta H_f$ respectively, as illustrated in Figure 26. $C_T$ is the Turnbull coefficient and $N_A$ Avogadro's number

Figure 25 reveals that, within experimental error, the reduced interfacial energies, inferred from nucleation investigations, of all charged sphere samples show a linear increase for increasing meta-stability. This suggests the use of the scheme sketched in Figure 26 and we simply extrapolate the data to zero $\Delta\mu$ without making any further assumptions. We obtain the extrapolated interfacial energy at equilibrium and the slope by performing least-square fits of $\sigma = \sigma_o + m\Delta\mu$.

The scheme according to Figure 27 is further used to estimate the molar entropy of fusion [44]. It is assumed that $\Delta S_f$ does not change with increasing metastability, i.e., with increasing particle number density. This has been shown to apply for hard sphere systems [182] and further has been observed for many metallic systems [183]. Both are systems where the hard-core repulsion creates an excluded volume, which is dominating the behavior of the condensed phase. In the present systems with their electrostatic interaction, the interaction is much softer, but still the repulsive part dominates the observed ordering processes. For further procedure Turnbull's rule which was found for metals is applied for the colloidal systems: $\sigma = C_T \Delta H_f / N_A$, where $N_A$ is Avogadro's number. At equilibrium $\Delta\mu = 0$, $\Delta H_f / N_A = T_M \Delta S_f / N_A$. With $\Delta S_f / N_A = $ const and $\sigma = C_T \Delta H_f / N_A$ this implies that at $\sigma = 0$, $\Delta H_f / N_A = 0$ and $\Delta\mu = -T_M \Delta S_f / N_A$, where the melting temperature is identified with the ambient temperature $T_M = 298$ K. Therefore, extrapolating the data to the intercept with the $\Delta\mu$ axis yields an estimate of the entropy of fusion $\Delta S_f$ and the enthalpy of fusion $\Delta H_f$ at equilibrium. Finally, the slope of the curve $m = \sigma_o/(T_M \Delta S_f / N_A)$ corresponds to Turnbull's coefficient $m = C_T$. Values for $m$ range between 0.235 and 0.405, each with small statistical uncertainties reflecting the good linear correlation of $\sigma$ and $\Delta\mu$. The spread of values is smaller than that observed for $\sigma_o$ and no clear correlation between $\sigma_o$ and $C_{T,bcc}$ is found. Moreover, none of the tests for correlations of $C_{T,bcc}$ to quantities relevant for the interaction strength and range was positive. The values of the molar entropy of fusion $\Delta S_f$ range between 1.5 and 4.6 J mol$^{-1}$ K$^{-1}$ and those for $\Delta H_f$ range between 0.45 and 1.36 kJ mol$^{-1}$. The values for molar $\Delta H_f$ are just below those of alkaline metals, which range about 2–3 kJmol$^{-1}$, but smaller than the values of about 8 kJmol$^{-1}$ for alkaline earth metals and 35.2 kJmol$^{-1}$ for tungsten [184]. Values for $\Delta S_f$ compare to a value of about 10 Jmol$^{-1}$ K$^{-1}$ for metals. For mono-disperse hard sphere systems, density functional theory calculations yield 9.7 Jmol$^{-1}$ K$^{-1}$ [182].

An interesting further observation can be made for colloids, which cannot be made in atomic systems. Neglecting isotopy, atoms are monodisperse. Due to their intrinsic polydispersity, the individual colloidal spheres in a sample, however, show slight size differences. With an increasing amount of polydispersity, also the entropy of a given phase increases due to additional compositional degrees of freedom. This effect is different for crystal and melt phase. It is smaller in the melt, since only a short-range order is affected. Therefore, if the size polydispersity increases, the entropy of fusion, i.e. the entropy difference between melt and solid, decreases. This leads to an interesting, somewhat counter-intuitive consequence of Turnbull' rule, namely that the interfacial fee energy between melt and solid also decreases with increasing polydispersity. In fact, our recent study could demonstrate, that the extrapolated equilibrium interfacial free energies showed a clear anti-correlation with the polydispersity, while the above derived entropies of fusions were line-arly correlated to the polydispersity. Therefore, spherical particles with slightly different diameters nucleate much faster than their strictly monodisperse counterparts in theory and simulation.

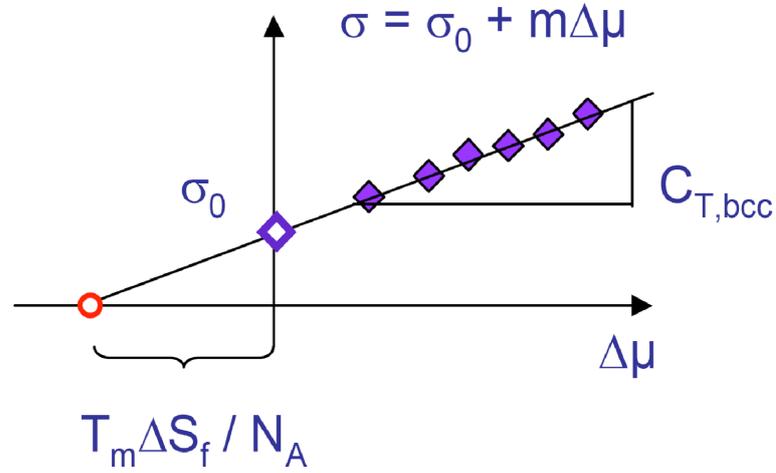

**Figure 27:** Extrapolation scheme based on Turnbull's rule: $\sigma_o = C_T \Delta H_f /N_A = C_T T_M \Delta S_f /N_A$. with $\Delta H_f = T_M \Delta S_f$. (Reproduced from T. Palberg, P. Wette, D.M. Herlach, *Equilibrium fluid-crystal interfacial free energy of bcc-crystallizing aqueus suspensions of polydisperse charged spheres*, Physical Review E **93** (2016) 022601, with the permission of APS Publishing, copyright 2016 APS).

### 3.2.4 Turnbull plots for colloids, metals and simulations: a comparison

So far we have presented results on nucleation rate densities and interfacial energies determined for fcc crystallizing metals and bcc crystallizing colloids. This now allows constructing Turnbull plots for both physically different systems and to compare them to theoretical predictions available for both structures [185, 186, 187]. Turnbull's rule was formulated using the equilibrium interfacial free energy. However, the linear dependence of the normalized interfacial energies on the meta-stability $\Delta \mu$ in Figure 27 shows that the very same Turnbull coefficients also apply in the meta-stable state of the liquid phase.

In Figure 28 the colloidal Turnbull coefficient $C_{T,bcc}$ is plotted versus the reduced equilibrium interfacial energies (in eV per atom) against the enthalpy of fusion (in eV per atom). For comparison, we also show the currently available data on other systems from the literature. Experimental investigations have been performed for bcc crystallizing metals yielding a Turnbull coefficient $C_{T,fcc} = 0.43$ [171], while simulation results for fcc crystallizing metals are better described by $C_{T,fcc} = 0.55$ [188]. The few simulations available for bcc-crystallizing metals are best described by $C_{T,bcc} = 0.29$ [188]. From an error-weighted linear fit to the data onn bcc crystallizing colloids an averaged Turnbull coefficient of $C_{T,bcc,expt} = 0.31 \pm 0.03$ is obtained. Using only the low-uncertainty data with $\Delta \mu$ derived using Würth's approximation for $\Delta \mu$, a slightly lower value of $C_{T,bcc,expt} = 0.25 \pm 0.02$ is achieved. Our values appear to be remarkably close to those from simulations of bcc metals (open triangles [188]) which yield an average $C_{T,bbc,sim} = 0.29$ (dash-dotted line) but are much smaller than the values for fcc crystallizing systems. These results further support the theoretical predictions based on entropic considerations that $C_{T,bcc}$ should be considerably smaller than $C_{T,fcc}$ [185, 186, 187].

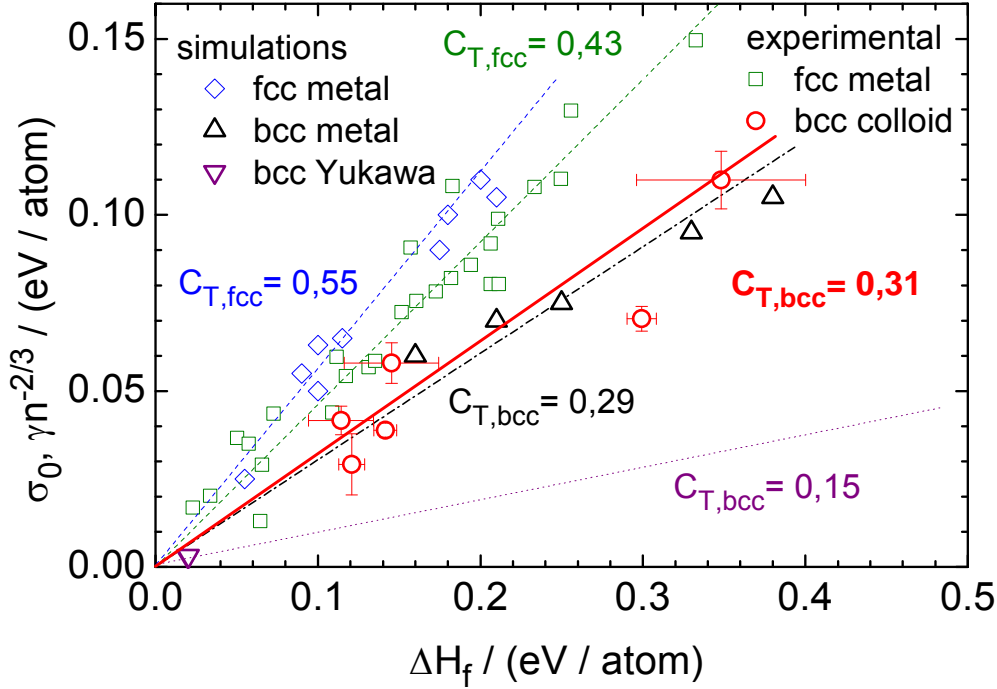

**Figure 28:** Turnbull plot of the reduced interfacial free energy versus the enthalpy of fusion. For better comparison, the data are here given in units of eV. Shown are data for bcc-crystallizing colloids (circles), simulation data for bcc-crystallizing metals (up-triangles) [188], bcc-crystallizing point Yukawa systems (down triangles) [189], and fcc crystallizing metals (diamonds) [188] as well as experimental data for fcc crystallizing metals (squares) [171]. Lines correspond to the indicated average values of Turnbull coefficients as quoted from [189, 188] and [171] for simulation and experimental results, respectively. An average value of $C_{T,bcc} = 0.31 \pm 0.03$ (thick solid line) is found for the bcc crystallizing colloids which appears to be remarkably close to that expected for bcc metals. (Reproduced from T. Palberg, P. Wette, D.M. Herlach, *Equilibrium fluid-crystal interfacial free energy of bcc-crystallizing aqueus suspensions of polydisperse charged spheres*, Physical Review E **93** (2016) 022601, with the permission of APS Publishing, Copyright 2016 APS).

## 4. Conclusion

Containerless processing by electromagnetic and electrostatic levitation has been introduced as powerful experimental tools to investigate short-range ordering and nucleation in undercooled metallic melts. Exemplary results are presented on Zr, quasicrystal forming alloys and Zr-based glass forming alloys. A statistical analysis of the distribution function of maximum undercooling in electrostatic levitation experiments hints on the onset of homogeneous nucleation in undercooled Zr-melts. From the homogeneous nucleation rate calculated within classical nucleation theory a lower limit of the solid-liquid interfacial energy was deduced. This value indicates that results by density functional theory and mo-lecular dynamics simulations may lead to an underestimation of the solid-liquid interfacial energies.

The formation of critical nuclei in undercooled metals takes place by atomic attachment frequencies in the order of the Debye frequency of $10^{13}$ Hz. Moreover, metals are not transparent for optical observations. Therefore, in situ investigations of crystal nucleation is extremely difficult for metallic systems if not all impossible. Nowadays, it is well known that colloidal suspensions are discussed as model systems to study crystal nucleation. Colloids are transparent for visible light at least at small particle concentrations and the particle kinetics are by several orders of magnitude more sluggish than atomistic and molecular systems. We introduce experimental tools like optical microscopy, light scattering and Ultra Small Angle X-ray Scattering to in situ investigate crystal nucleation in colloidal suspensions. Since homogeneous and heterogeneous nucleation is easily discriminated in the colloids, important parameters for crystal nucleation like the interfacial energy are directly inferred from investigations of homogeneous nucleation. Results are presented for a variety of different colloidal suspensions. With these results it is demonstrated that even Turnbull coefficients can be determined. They are compared with results of experimental investigations in pure metals and simulations and, thereby, confirm existing predictions on the structural dependence of Turnbull's coefficients. Our detailed and comparative studies strongly support the idea that indeed colloidal suspensions are suitable model systems for nucleation phenomena also in metallic systems.

*Acknowledgements*

The authors are very grateful to many co-workers and colleagues for careful experimental investigations inevitable for the scientific results presented in this work. In particular we thank Dirk Holland-Moritz for his valuable contributions to experimental investigations of short range order in both metals and colloids and nucleation studies, Patrick Wette for his experimental work and analysis of light scattering and USAXS investigations on colloids. These experiments on metals and colloids were further supported by Wolfgang Hornfeck, Nina Lorenz just to mention some. We are also very much indebted to many colleagues for continued stimulating discussions as Hans-Joachim Schöpe, Peter Galenko, J. Horbach, and others. The work was supported by the Deutsche Forschungsgemeinschaft within the projects located in Mainz, Grants No. Pa459/16, No. Pa459/17, and those located at DLR Köln, Grants No. HE1601/17, HE1601/21, He1601/24). One of the author (DMH) is also grateful for financial support to German Aerospace Center Space Management within contract 50WM1140, and the European Space Agency within contract 15236/02/NL/SH.